\documentclass{aa}  

\bibpunct{(}{)}{;}{a}{}{,}

\usepackage{graphicx, graphics}
\usepackage{txfonts}
\usepackage{longtable}

\begin{document} 

   \title{Physical properties of accretion shocks toward the Class I protostellar system Oph-IRS~44}
   
   \author{
   E. Artur de la Villarmois \inst{1,2}
   \and V. V. Guzm\'an \inst{1,2}
   \and J. K. J{\o}rgensen \inst{3}
   \and L. E. Kristensen \inst{3}
   \and E. A. Bergin \inst{4}
   \and D. Harsono \inst{5}
   \and N. Sakai \inst{6}
   \and E. F. van Dishoeck \inst{7,8}
   \and S. Yamamoto \inst{9}
   }

   \institute{Instituto de Astrof\'isica, Pontificia Universidad Cat\'olica de Chile, Av. Vicu\~na Machenna 4860, 7820436 Macul, Santiago, Chile \\
   \email{eartur@astro.puc.cl}
   \and N\'ucleo Milenio de Formaci\'on Planetaria -- NPF, Universidad de Valpara\'iso, Av. Gran Breta\~na 1111, Valpara\'iso, Chile
   \and Niels Bohr Institute, University of Copenhagen, {\O}ster Voldgade 5--7, 1350 Copenhagen K., Denmark
   \and Department of Astronomy, University of Michigan, 311 West Hall, 1085 S. University Ave, Ann Arbor, MI 48109, USA
   \and Institute of Astronomy, Department of Physics, National Tsing Hua University, Hsinchu, Taiwan
   \and RIKEN Cluster for Pioneering Research, 2-1 Hirosawa, Wako-shi, Saitama 351-0198, Japan
   \and Leiden Observatory, Leiden University, PO Box 9513, NL-2300 RA Leiden, the Netherland
   \and Max-Planck Institut f$\ddot{\mathrm{u}}$r extraterrestrische Physik, Giessenbachstra{\ss}e 1, 85748, Garching bei M$\ddot{\mathrm{u}}$nchen, Germany
   \and Department of Physics, The University of Tokyo, Bunkyo-ku, Tokyo 113-0033, Japan
   }

\abstract
   {The final outcome and chemical composition of a planetary system depend on its formation history: the physical processes that were involved and the molecular species available at different stages. Physical processes such as accretion shocks are thought to be common in the protostellar phase, where the envelope component is still present, and they can release molecules from the dust to the gas phase, altering the original chemical composition of the disk. Consequently, the study of accretion shocks is essential for a better understanding of the physical processes at disk scales and their chemical output.}
   {The purpose of this work is to assess how the material from the infalling envelope feeds the disk and the chemical consequences thereof, particularly the characteristics of accretion shocks traced by sulfur-related species.}
   {We present high angular resolution observations (0.1", corresponding to 14~au)  with the Atacama Large Millimeter/submillimeter Array (ALMA) of the Class I protostar Oph-IRS 44 (also known as YLW 16A). The continuum emission at 0.87~mm is observed, together with sulfur-related species such as SO, SO$_{2}$, and $^{34}$SO$_{2}$. The non-local thermodynamic equilibrium (non-LTE) radiative-transfer tool RADEX and the rotational diagram method are employed to assess the physical conditions of the SO$_{2}$ emitting region.}
   {Six lines of SO$_{2}$, two lines of $^{34}$SO$_{2}$, and one line of SO are detected toward IRS 44. The emission of all the detected lines peaks at $\sim$0.1" ($\sim$14~au) from the continuum peak and we find infalling-rotating motions inside 30 au. However, only redshifted emission is seen between 50 and 30~au. Colder and more quiescent material is seen toward an offset region located at a distance of $\sim$400~au from the protostar, and we do not find evidence of a Keplerian profile in these data. The SO$_{2}$ emitting region around the protostar is consistent with dense gas ($\geq$~10$^{8}$~cm$^{-3}$), temperatures above 70~K, high SO$_{2}$ column densities between 0.4 and 1.8~$\times$~10$^{17}$~cm$^{-2}$, line widths between 12 and 14~km~s$^{-1}$, and an abundance ratio SO$_{2}$/SO~$\geq$~1, suggesting that some physical mechanism is enhancing the gas-phase SO$_{2}$ abundance.}  
   {Accretion shocks are the most plausible explanation for the high temperatures, high densities, and velocities found for the SO$_{2}$ emission. The offset region seems to be part of a localized streamer that is injecting material to the disk--envelope system through a protrusion observed only in redshifted emission and associated with the highest kinetic temperature. When material enters the disk--envelope system, it generates accretion shocks that increase the dust temperature and desorb SO$_{2}$ molecules from dust grains. High-energy SO$_{2}$ transitions (\textit{E$_\mathrm{up}$}~$\sim$~200~K) seem to be the best tracers of accretion shocks that can be followed up by future higher angular resolution ALMA observations and compared to other species to assess their importance in releasing molecules from the dust to the gas phase.}
   
   
   \maketitle
   
   \section{Introduction}
   
The formation and evolution of protoplanetary disks are fundamental in the process of low-mass star formation, such as the formation of our own Solar System. A typical low-mass star forms when a molecular cloud with angular momentum collapses, and a protostar is formed at the central part with an infalling-rotating envelope whose inner part evolves to a circumstellar disk \citep{Terebey1984, Shu1993, Hartmann1998}. Eventually, the star reaches its final mass, the envelope dissipates, and planets form in the disk. As a consequence, the final composition of planets is strongly dependent on the physical and chemical processes within the circumstellar disk. However, as disks first arise in the early stages of young stars \citep{Jorgensen2009, Harsono2014, Yen2015} and the first steps of planet formation may occur when they are still deeply embedded \citep[e.g.,][]{Harsono2018, Tychoniec2020}, the chemical evolution of the material as it is accreted from the infalling envelope may play a key role. 

The process of low-mass star formation comprises different stages \citep{Robitaille2006}, and Class I sources link the deeply embedded Class 0 sources (where the envelope is the dominant mass component) with the emergence of Class II disks (Keplerian disks with a negligible envelope). Class I sources are therefore the perfect candidates to study the connection between the envelope and the disk and, additionally, to investigate the dynamics and chemical composition of the young disk. 

Theoretical models predict that the material from the envelope falls on the circumstellar disk and produces accretion shocks at the envelope--disk interface \citep{Stahler1994, Yorke1999, Krasnopolsky2002}. These accretion shocks have been invoked to explain the observed jump in density and drastic enhancement of SO toward the Class 0 and I sources L1527 and TMC-1A \citep{Sakai2014, Sakai2016}, the asymmetric accretion found toward TMC-1A \citep{Hanawa2022}, and the emission of SO and SO$_{2}$ at the edge of the disks from two Class I/II sources, DG Tau and HL Tau \citep{Garufi2022}. Accretion shocks in dense ($\geq$~10$^{8}$~cm$^{-3}$) gas induce an increase in the dust temperature, and species that are locked in grain mantles are subsequently released into the gas phase, which affects the chemical content of the early disk \citep{vanGelder2021}. Although the presence of accretion shocks explains the jump in abundances observed for shock-related species and is the most plausible mechanism deduced from numerical simulations \citep{Miura2017}, only a few low-mass protostars show evidence of accretion shocks to date \citep[e.g.,][]{Lee2014, Sakai2014, Garufi2022}, and their physical parameters are not well constrained observationally. Apart from accretion shocks, contributions from disk winds or outflows would also be important \citep[e.g.,][]{Bjerkeli2016, Alves2017, Tabone2017, Harsono2021}. Therefore, observations at disk scales ($\sim$100~au) need to be performed to confirm the existence of accretion shocks, understand the origin of the observed abundances, and assess the physical parameters associated with this mechanism. 

A suitable source for proving the nature of accretion shocks is Oph-IRS~44, a Class I source located in the Ophiuchus molecular cloud at a distance of 139~pc \citep[average value for the L1688 cloud;][]{Canovas2019}. \cite{Artur2019} detected strong SO$_{2}$ emission toward a compact region ($\leq$~60~au) in IRS~44, with an angular resolution of 0$\farcs$4 ($\sim$60~au). This particular SO$_{2}$ transition (18$_{4,14}$ -- 18$_{3,15}$) is associated with an upper-level energy (\textit{E$_\mathrm{up}$}) of $\sim$200~K and its line profile shows a velocity range of $\sim$20~km~s$^{-1}$. The angular resolution of 0$\farcs$4 of the data was not high enough to resolve the SO$_{2}$ emission and provide strong conclusions for the possible origin scenarios: accretions shocks, disk winds, or outflows.

IRS~44 was first identified as YLW 16A by \cite{Young1986} through \textit{IRAS} observations, and other common names are Oph-emb~13, ISO-Oph 143, LFAM 35, and [GY92]~269, among others. It is associated with a bolometric temperature (\textit{T$_\mathrm{bol}$}) of 280~K, a bolometric luminosity (\textit{L$_\mathrm{bol}$}) of 7.1~L$_{\odot}$ \citep{Evans2009}, and an envelope mass (\textit{M$_\mathrm{env}$}) of 0.051~M$_{\odot}$ \citep[for a distance of 139~pc;][]{Jorgensen2009}. IRS~44 has been proposed to be a protobinary system with a separation of $\sim$0$\farcs$3, based on observations with the \textit{Hubble Space Telescope} \citep[\textit{HST};][]{Allen2002}, the Very Large Telescope \citep[VLT;][]{Duchene2007}, and the \textit{Spitzer Space Telescope} \citep{McClure2010}. Nevertheless, there is no evidence of a binary component in the submillimeter  regime, through ALMA band 6 and band 7 observations \citep{Sadavoy2019, Artur2019}. 

In this paper we present high angular resolution 0$\farcs$1 (14~au) ALMA observations of multiple SO$_{2}$ molecular lines toward IRS 44. We discuss their potential to trace accretions shocks, and provide values of the physical parameters for the emitting gas. Section 2 describes the observational procedure, calibration, and the parameters of the observed molecular transitions. The observational results are presented in Sect. 3, while Sect. 4 is dedicated to the analysis of the data, with position-velocity diagrams, radiative-transfer models, estimations of rotational and excitation temperatures, and calculations of molecular column densities. We discuss the structure and kinematics of IRS~44 in Sect. 5, and end with a summary in Sect. 6.

\section{Observations}

IRS 44 was observed with ALMA during 2021 May 17 and 18 as part of the program 2019.1.00362.S (PI: Elizabeth Artur de la Villarmois). At the time of the observations, 47 and 45 antennas were available, respectively, in the array providing baselines between 15 and 2517~m. The observations targeted nine different spectral windows to observe multiple SO$_{2}$ lines, the less abundant $^{34}$SO$_{2}$ isotopolog, and SO. The observed molecular transitions and their spectroscopic data are summarized in Table~\ref{table:observations}. 

The calibration and imaging were done in CASA\footnote{http://casa.nrao.edu/} version 6.1.1 (McMullin et al. 2007). Gain and bandpass calibrations were performed through the observation of the quasars J1517–2422 and J1700–2610. Imaging was performed using the \texttt{tclean} task in CASA, where the Briggs weighting with a robust parameter of 0.5 was employed. The automasking option was chosen and the channel resolution is 0.21~km~s$^{-1}$. The resulting dataset has a beam size of 0$\farcs$13~$\times$~0$\farcs$09 (18~$\times$~13~au) with a position angle (PA; measured from north to east) of -81$\degr$ and a largest angular scale (LAS) of 2$\farcs$3. The continuum rms level is 0.08~mJy beam$^{-1}$ and the rms level of each spectral window is listed in Table~\ref{table:observations}.

\begin{table*}[ht]
        \caption{Spectral setup and parameters of the observed molecular transitions.}
        \label{table:observations}
        \centering
        \begin{tabular}{l l c r c c}
                \hline\hline
                Species                	 	& Transition                           	& Frequency 	& \textit{E$_\mathrm{up}$} 	& \textit{A$_{ij}$}                         	& rms $^{(a)}$          		\\
                                              		&                                          	& [GHz]        	&  [K]                                  	& [$\times$10$^{-5}$~s$^{-1}$]       	& [mJy beam$^{-1}$    	\\
                                              		&                                             	&                    	&                                            	&                                                     	&  per channel]               	\\
                \hline
                SO                               	& 10$_{11}$ -- 10$_{10}$      	& 336.5538    	& 143                   			& \enspace 1                                   	& 2.8                              	\\
                SO$_{2}$                    	& 16$_{7,9}$ -- 17$_{6,12}$  	& 336.6696    	& 245                   			& \enspace 6                                 	& 2.8                                	\\
                SO$_{2}$                   	& 18$_{4,14}$ -- 18$_{3,15}$  	& 338.3060    	& 197                   			& 33                                                	& 2.4                                	\\
                SO$_{2}$               		& 20$_{1,19}$ -- 19$_{2,18}$  	& 338.6118    	& 199                  			& 29                                                	& 2.5                              	\\
                SO$_{2}$    			& 24$_{2,22}$ -- 23$_{3,21}$  	& 348.3878    	& 293                  			& 19                                           		& 2.7                               	\\
                SO$_{2}$              		& 5$_{3,3}$ -- 4$_{2,2}$        	& 351.2572    	& 36                           		& 34                                                	& 3.4                              	\\
                SO$_{2}$                          	& 10$_{6,4}$ -- 11$_{5,7}$   	& 350.8628    	& 139                   			& \enspace 4                                  	& 3.2                               	\\
                $^{34}$SO$_{2}$         	& 14$_{4,10}$ -- 14$_{3,11}$ 	& 338.7857   	& 134                   			& 31                                                	& 2.5                              	\\
                $^{34}$SO$_{2}$           	& 19$_{4,16}$ -- 19$_{3,17}$ 	& 348.1175   	& 213                   			& 35                                                	& 2.8                               	\\
                $^{34}$SO$_{2}$$^{(b)}$	& 9$_{6,4}$ -- 10$_{5,5}$       	& 351.0896    	& 127                   			& \enspace 4                                    	& 2.8                             	\\
                \hline\hline
        \end{tabular}
        \tablefoot{\textit{E$_\mathrm{up}$} and  \textit{A$_{ij}$} correspond to the upper-level energy and the Einstein  \textit{A} coefficient, respectively. $^{(a)}$The channel width is 0.21~km~s$^{-1}$. $^{(b)}$ Not detected.}
\end{table*}

\section{Results}
\subsection{Continuum emission}

\begin{figure}[h]
        \centering
        \includegraphics[width=.49\textwidth]{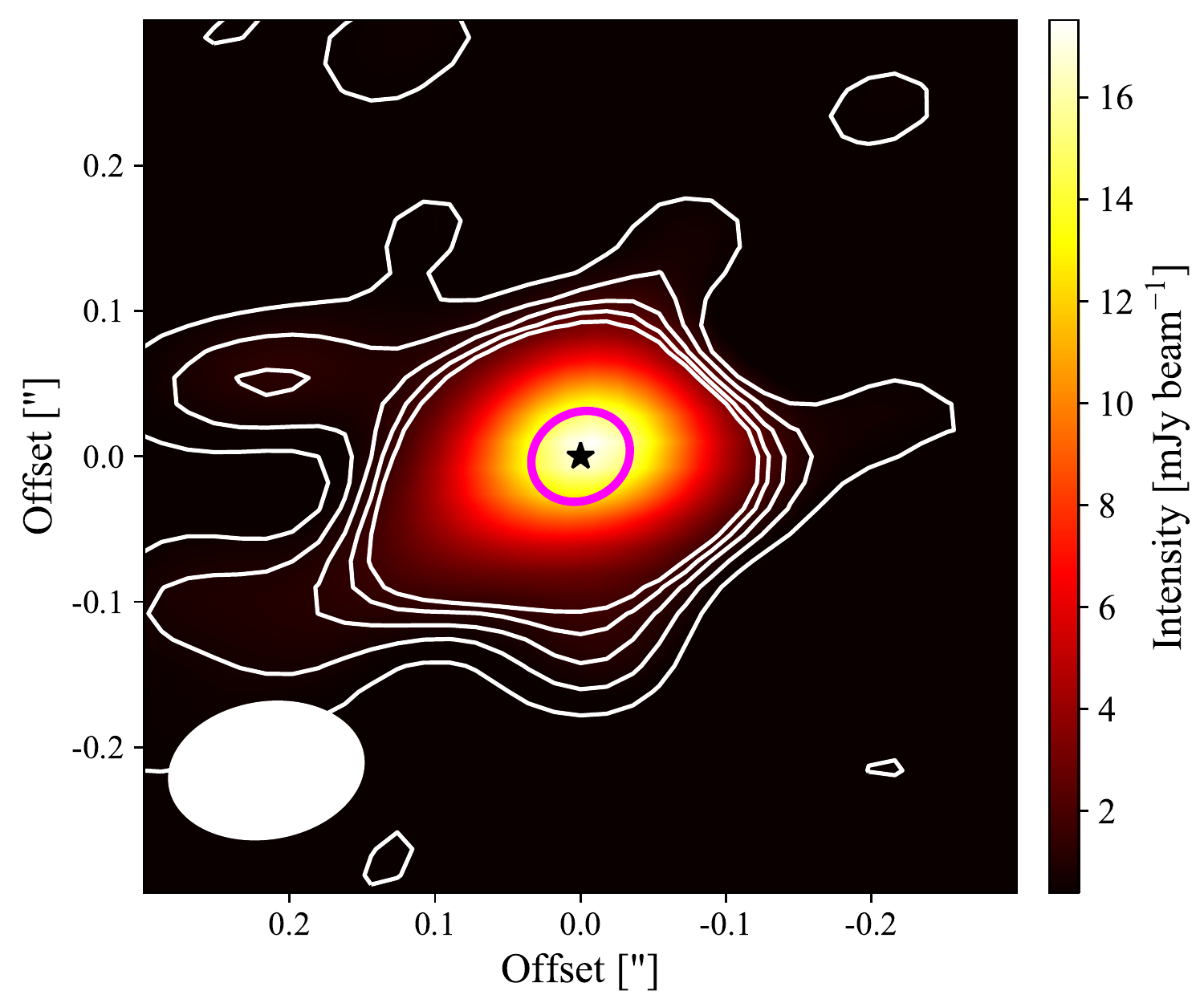}
        \caption[]{\label{fig:continuum}
        Continuum emission (0.87~mm) toward IRS~44 above 5$\sigma$ ($\sigma$~=~0.08~mJy~beam$^{-1}$). The white contours represent the weakest emission of [5$\sigma$, 10$\sigma$, 15$\sigma$, 20$\sigma$, and 25$\sigma$] for clarity. The black star shows the position of the continuum peak and the synthesized beam is indicated by the white filled ellipse. The magenta ellipse represents the deconvolved size from the 2D Gaussian fit.
        }
\end{figure}

The continuum emission is shown in Fig.~\ref{fig:continuum}, where the horizontal component is slightly more extended than the vertical component, and the emission above 5$\sigma$ is contained within a radius of 0$\farcs$2 ($\sim$30~au). Two-dimensional (2D) Gaussians are used to fit emission in the image plane, obtaining an integrated flux of 22.9~$\pm$~1.0~mJy, a peak flux of 16.91~$\pm$~0.48~mJy~beam$^{-1}$, and a deconvolved size of (0$\farcs$07~$\pm$~0$\farcs$01)~$\times$~(0$\farcs$06~$\pm$~0$\farcs$01) with a PA of 119~$\pm$~74$\degr$ (see the magenta ellipse in Fig.~\ref{fig:continuum}). The continuum peak position corresponds to $\alpha$~=~16$^\mathrm{h}$27$^\mathrm{m}$27$^\mathrm{s}$.9858~$\pm$~0$^\mathrm{s}$.0002 and $\delta$~=~-24$\degr$39$\arcmin$34$\farcs$063~$\pm$~0$\farcs$001.

The disk mass at 0.87~mm was calculated from the continuum flux (22.9~$\pm$~1.0~mJy) and using Eq. (2) from \cite{Artur2018}, which assumed optically thin emission, an opacity of 0.0175~cm$^{-2}$ per gram of gas at 0.87~mm, and a dust temperature of 30~K. A total mass \textit{M$_\mathrm{gas + dust}$} of (4.0~$\pm$~0.2)~$\times$~10$^{-3}$~M$_{\odot}$ was obtained, adopting a dust temperature (\textit{T$_\mathrm{dust}$}) of 15~K, the value proposed by \cite{Dunham2014b} for Class I sources. If \textit{T$_\mathrm{dust}$}~=~30~K is assumed, the total mass   decreases by a factor of $\sim$3. Given that the dust emission at 0.87~mm could be optically thick toward a Class I source, the calculated \textit{M$_\mathrm{gas + dust}$} represents a lower limit for the total mass.

\cite{Allen2002} and \cite{Duchene2007} suggested that IRS~44 is a protobinary system with a separation of $\sim$0$\farcs$3. However, the continuum emission at 0.87~mm shows no binary detection in our ALMA data and we can only set an upper limit of 7~$\times$~10$^{-5}$~M$_{\odot}$ for the total mass of a possible binary component (for a value of 5$\sigma$ and adopting the same parameters as in the previous paragraph).

\begin{figure*}[h!]
        \centering
        \includegraphics[width=.29\textwidth]{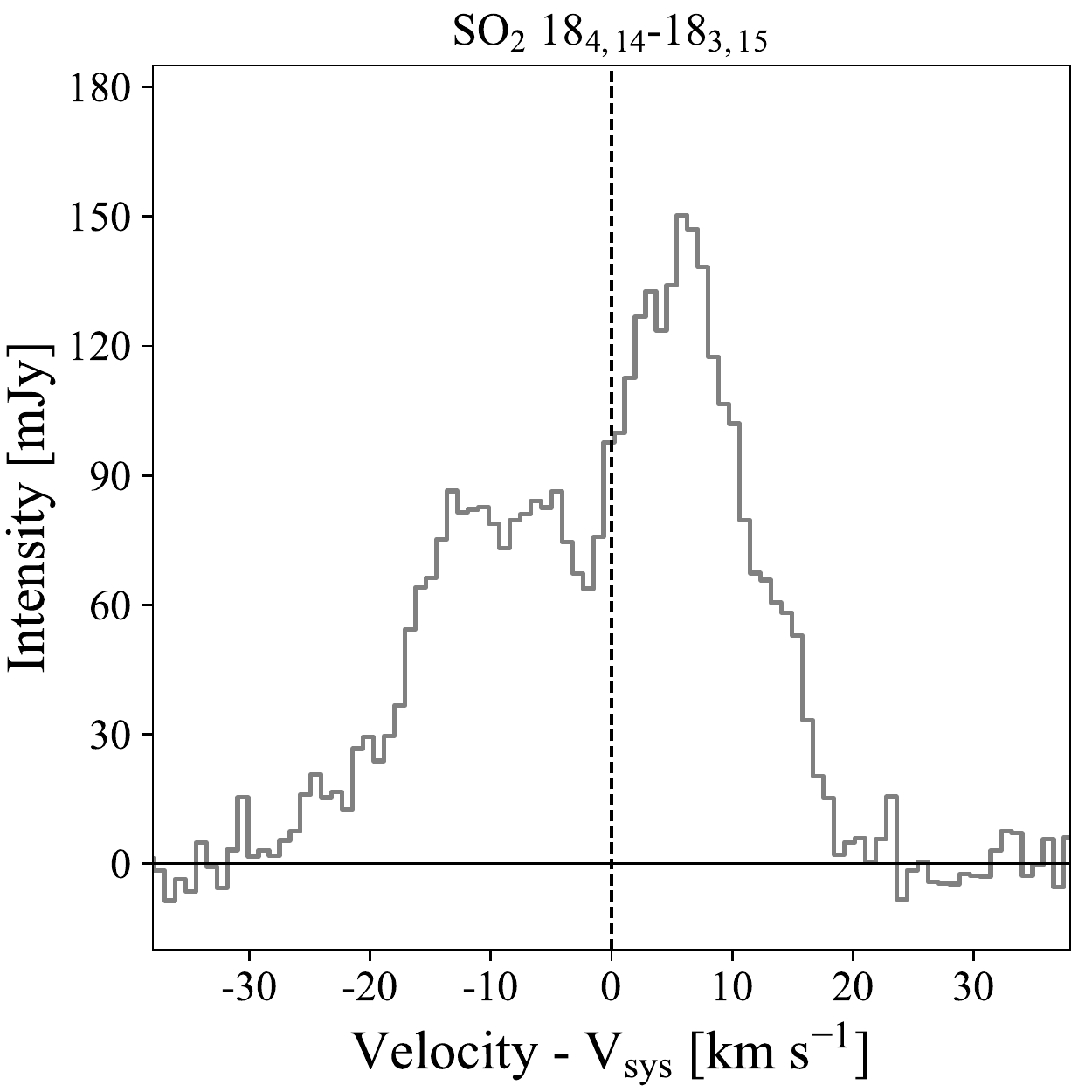}
        \includegraphics[width=.335\textwidth]{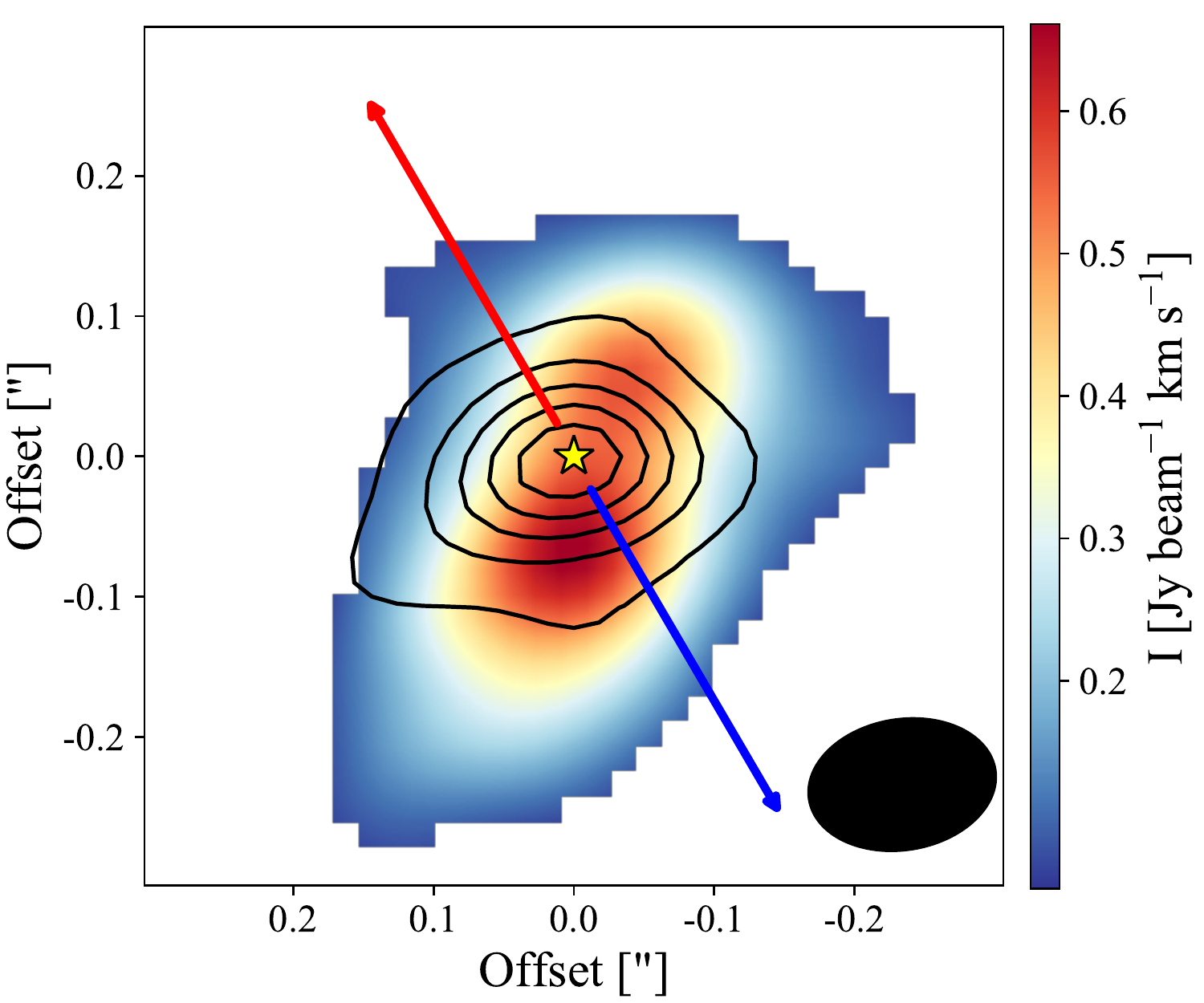}
        \includegraphics[width=.335\textwidth]{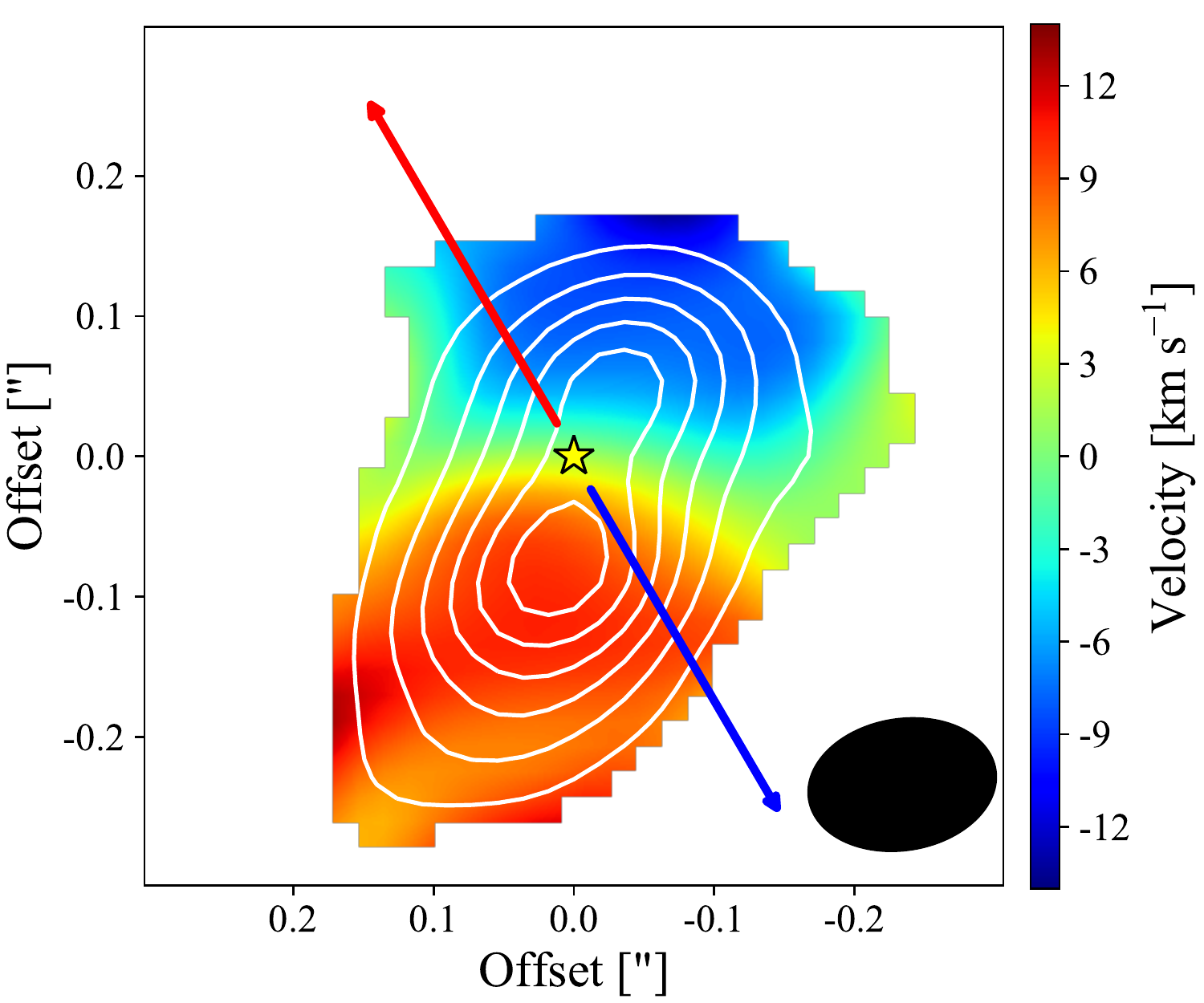}
        \caption[]{\label{fig:central}
        Emission of the SO$_{2}$ 18$_{4,14}$ -- 18$_{3,15}$ line. \textit{Left:} Spectrum rebinned by a factor of 4, integrated over a circular region with \textit{r}~=~0$\farcs$2, and centered on the continuum peak position. \textit{Center:} Moment 0 map above 3$\sigma$ (color scale), integrated over 60~km~s$^{-1}$, and continuum emission (black contours), starting at 20$\sigma$ and following   steps of 40$\sigma$. \textit{Right:} Moment 1 map (color scale) and moment 0 map (white contours) above 3$\sigma$. The blue and red arrows show the direction of the outflow from \cite{vanderMarel2013}, the yellow star indicates the continuum peak position, and the synthesized beam is represented by the black filled ellipse in the bottom right corner. The adopted systemic velocity is 3.7~km~s$^{-1}$.
        }
\end{figure*}

\subsection{Molecular transitions}

All the targeted molecular lines listed in Table~\ref{table:observations} were detected toward IRS~44, with the exception of the $^{34}$SO$_{2}$ 9$_{6,4}$ -- 10$_{5,5}$ line. This nondetection is consistent with the low Einstein \textit{A} coefficient (\textit{A$_{ij}$}~=~4~$\times$~10$^{-5}$~s$^{-1}$) of the transition and it being a less abundant isotopolog. The six detected SO$_{2}$ lines have different upper level energies \textit{E$_\mathrm{up}$}, covering a broad range from 36 to 293~K. The brightest emission toward the continuum peak is from the SO$_{2}$ 18$_{4,14}$ -- 18$_{3,15}$ line with \textit{E$_\mathrm{up}$}~=~197~K, while there is an offset region located at a distance of $\sim$3$\farcs$0 ($\sim$400~au) from the protostar that shows bright emission of the SO$_{2}$ line related with the lowest energy: SO$_{2}$ 5$_{3,3}$ -- 4$_{2,2}$ with \textit{E$_\mathrm{up}$}~=~36~K. 

Figure~\ref{fig:central} presents the spectrum, and moment 0 and 1 maps of the SO$_{2}$ 18$_{4,14}$ -- 18$_{3,15}$ line toward the central region. The spectrum was taken over a circular region with \textit{r}~=~0$\farcs$2 and shows a  broad-line profile, from -20 to 20~km~s$^{-1}$, and a decrease in the emission around the systemic velocity (\textit{V$_\mathrm{sys}$}) of 3.7~km~s$^{-1}$, estimated from previous APEX observations \citep{Lindberg2017}. The moment 0 map reveals that the emission is concentrated around the protostar; however, the emission peak is  slightly offset from the continuum peak, $\sim$0$\farcs$1 ($\sim$14~au), and corresponds to the  redshifted component. The moment 1 map shows a clear rotational signature   from northwest to southeast, with a PA of 157~$\pm$~3$\degr$. The PA for the SO$_{2}$ 18$_{4,14}$ -- 18$_{3,15}$ line emission was   obtained from a 2D Gaussian fit of the moment 0 map. We note that this PA value is not perpendicular to the outflow direction (PA~=~20$\degr$), which was   estimated by \cite{vanderMarel2013} using single-dish observations of CO 3-2. For the other detected lines (five SO$_{2}$, two $^{34}$SO$_{2}$, and one SO line), the spectra and  moment 0 and 1 maps are presented in the Appendix, in Figs.~\ref{fig:spectra}, \ref{fig:mom}, and \ref{fig:mom_bis}, showing that SO$_{2}$, $^{34}$SO$_{2}$, and SO exhibit a similar nature: broad spectra, emission concentrated around the protostar, and a clear rotational signature. In addition, all the detected transitions show that the peak of emission is offset south from the continuum peak position, at a distance of $\sim$0$\farcs$1 ($\sim$14~au). On average, the six SO$_{2}$ transitions show a full width at half maximum (FWHM) value of 12~km~s$^{-1}$ for the blueshifted emission and 14~km~s$^{-1}$ for the redshifted emission. Integrated fluxes of the observed transitions are presented in Table~\ref{table:fluxes} in the Appendix.

\begin{figure*}[h!]
        \centering
        \includegraphics[width=.99\textwidth]{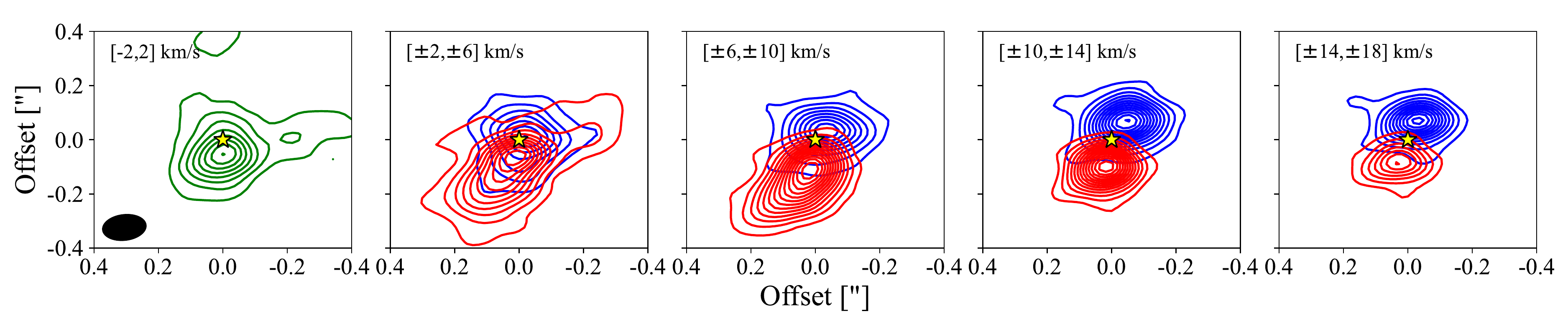}
        \caption[]{\label{fig:contour}
        Contour maps of SO$_{2}$ 18$_{4,14}$ -- 18$_{3,15}$ for different velocity ranges, shifted to 0 velocity. The contours start at 5$\sigma$ and follow  steps of 5$\sigma$. The panel that includes the systemic velocity (3.7~km~s$^{-1}$) is shown by the green contours in the first panel, while blue- and redshifted emission is represented by the blue and red contours, respectively. The yellow star shows the position of the source and the synthesized beam is indicated by the black filled ellipse in the bottom left corner of the first panel. 
        }
\end{figure*}

\begin{figure*}[h]
        \centering
        \includegraphics[width=.43\textwidth]{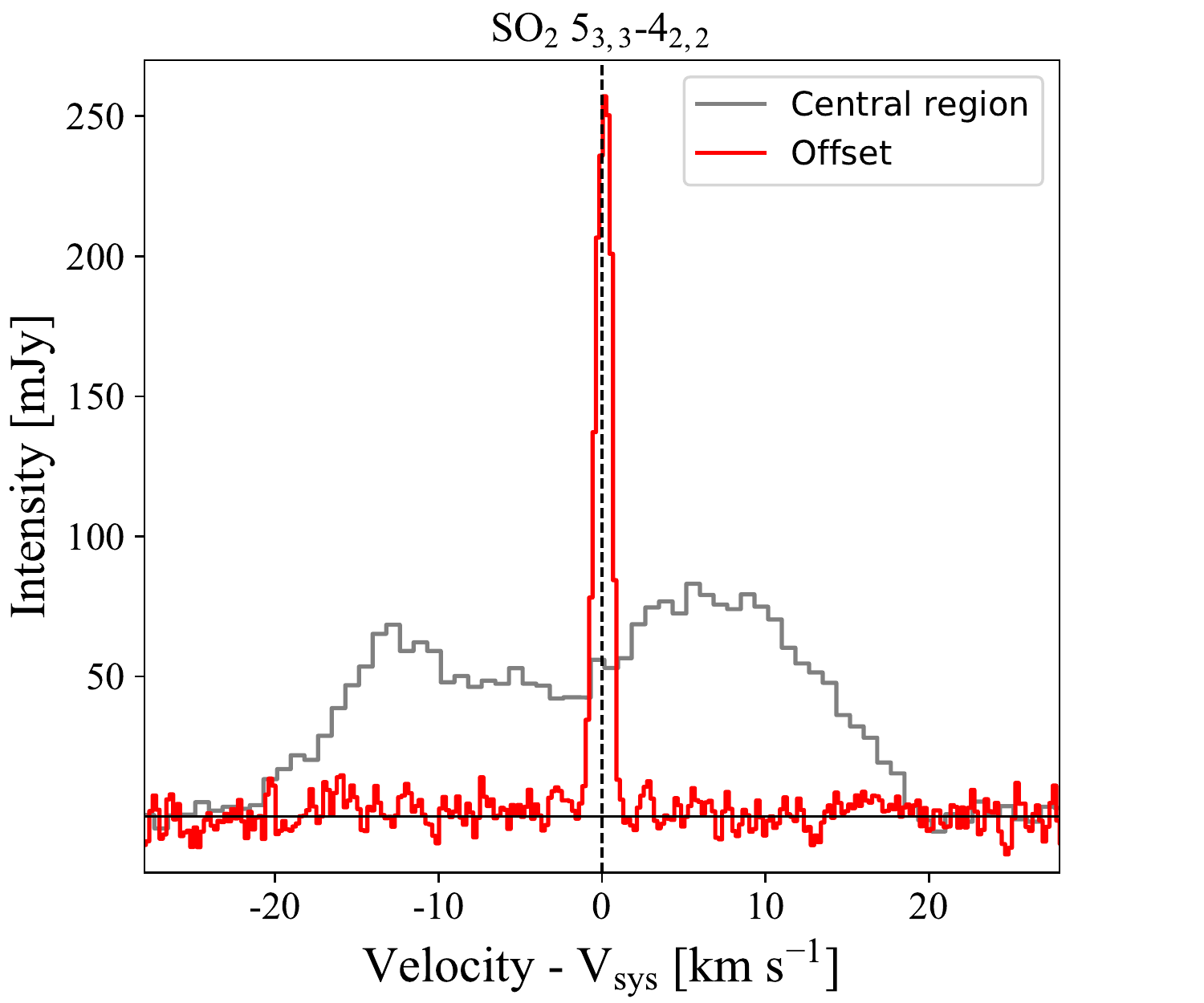}
        \includegraphics[width=.45\textwidth]{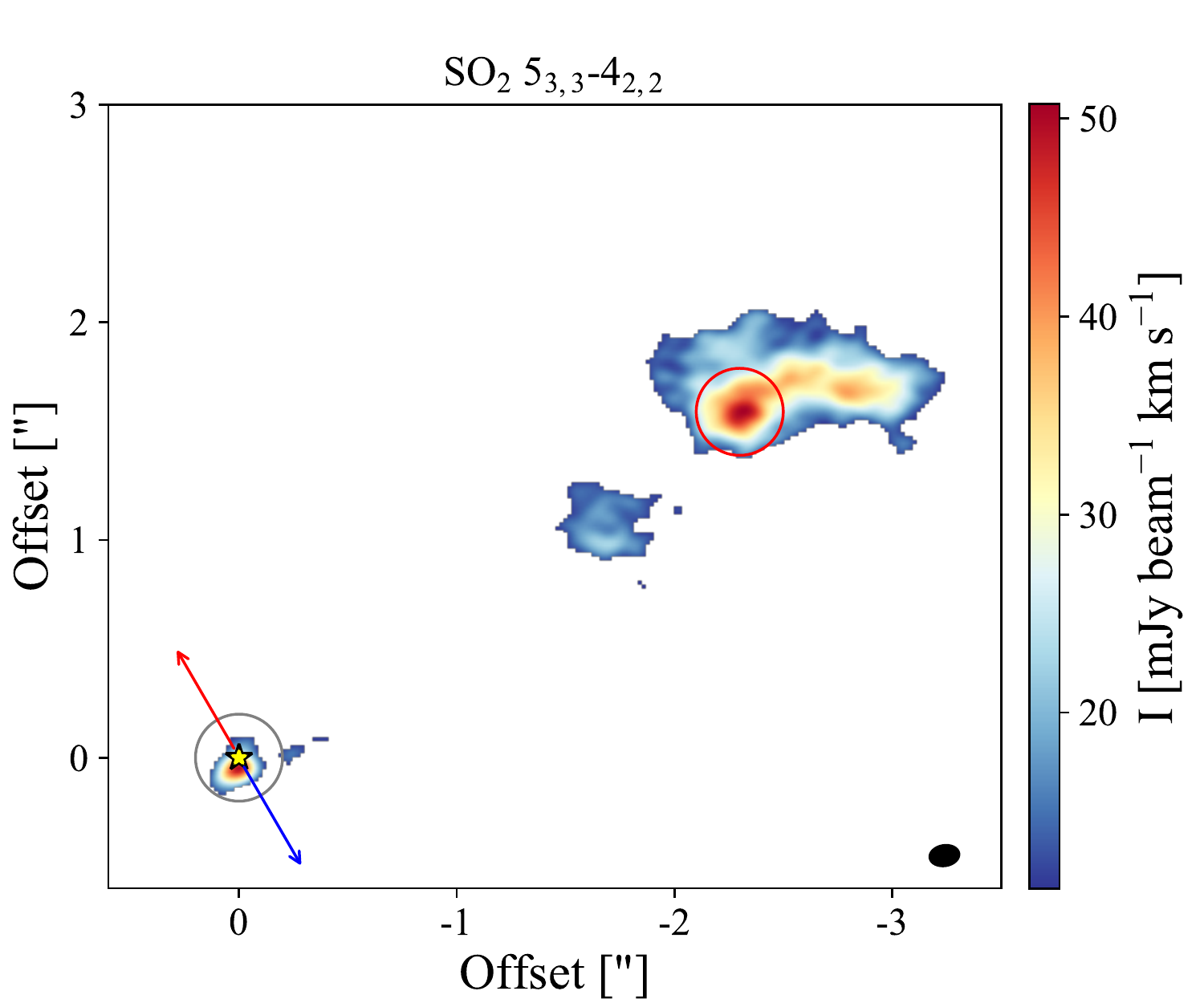}
        \caption[]{\label{fig:offset}
        Emission of the SO$_{2}$ 5$_{3,3}$ -- 4$_{2,2}$ line. \textit{Left:} Spectra integrated over a circular region with \textit{r}~=~0$\farcs$2, centered on the continuum peak position (gray) and centered on the offset region (red). The spectrum taken at the continuum peak position has been rebinned by a factor of 4. \textit{Right:} Moment 0 map above 3$\sigma$ integrated over 60~km~s$^{-1}$. The gray and red circles represent the regions from which the spectra in the left panel were taken. The blue and red arrows show the direction of the outflow, the yellow star indicates the continuum peak position, and the synthesized beam is represented by the black filled ellipse in the bottom right corner. The adopted systemic velocity is 3.7~km~s$^{-1}$.
        }
\end{figure*}

Figure~\ref{fig:contour} shows contour maps of the SO$_{2}$ 18$_{4,14}$ -- 18$_{3,15}$ line for different velocity ranges. Low-velocity contours (between -2 and 2~km~s$^{-1}$) are concentrated around the protostar, but the weakest contours also present emission toward the west. Intermediate velocities, between $\pm$6 and $\pm$10~km~s$^{-1}$, show that the redshifted emission is more extended than the blueshifted emission, possibly related with a protrusion from a localized streamer. Finally, a clear and symmetric rotating signature around the protostar is seen for high velocities ($\geq$10~km~s$^{-1}$), which will be referred to as a disk--envelope  structure.

At larger angular scales, the SO$_{2}$ 5$_{3,3}$ -- 4$_{2,2}$ line (\textit{E$_\mathrm{up}$}~=~36~K) shows bright emission toward an offset region, located at a distance of 2$\farcs$8 ($\sim$400~au) from the protostar, and its spectra and moment 0 map are presented in Fig.~\ref{fig:offset}. The spectra were taken over the central region (gray) and the offset region (red), revealing that the offset region is associated with low velocities ($\leq$2~km~s$^{-1}$), in contrast with the broad-line profile observed toward the central region, suggesting a different and more quiescent origin. Two other SO$_{2}$ lines (with \textit{E$_\mathrm{up}$} value of 197 and 199~K) show weaker emission toward the offset region and their moment 0 maps are presented in Fig.~\ref{fig:mom_large} in the Appendix. Given that the SO$_{2}$ line with the lowest \textit{E$_\mathrm{up}$} value (36~K) shows the brightest emission, the offset region is associated with colder gas, more consistent with a cloudlet or an SO$_{2}$ knot with a PA of 125~$\pm$~7$\degr$. In this case the PA value was calculated by projecting a line that connects the continuum peak with the brightest pixel of the offset region, and it is consistent with the direction of the weakest contours seen in the first two panels of Fig.~\ref{fig:contour}. The offset region is henceforth referred to as an SO$_{2}$ knot.

\section{Analysis}

\subsection{Position--velocity diagrams}

\begin{figure*}[h!]
        \centering
        \includegraphics[width=.99\textwidth]{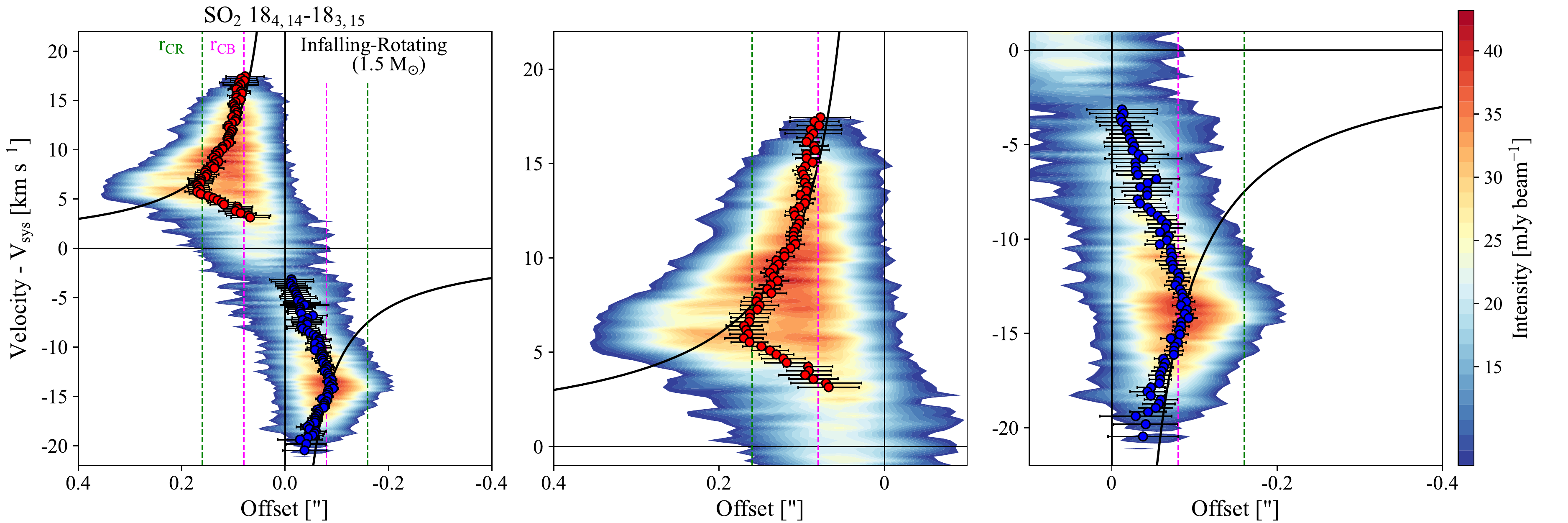}
        \caption[]{\label{fig:pv_25}
        Position--velocity diagram for SO$_{2}$ 18$_{4,14}$--18$_{3,15}$. \textit{Left}: Emission above 3$\sigma$, employing a PA of 157$^{\circ}$. The blue and red dots represent blue- and redshifted emission peaks above $\pm$3~km~s$^{-1}$, respectively. The black line represents an infalling-rotating profile with \textit{M$_{\star}$}~=~1.5M$_{\odot}$, an inclination of 70$^{\circ}$, and \textit{r$_\mathrm{CB}$}~=~0$\farcs$08 (magenta dashed lines). The \textit{r$_\mathrm{CR}$} is shown by the green dashed lies. \textit{Center}: Zoomed-in version for the redshifted emission. \textit{Right}: Zoomed-in version for the blueshifted emission. The systemic velocity is 3.7~km~s$^{-1}$.
         }
\end{figure*}

Figure~\ref{fig:pv_25} shows a position--velocity (PV) diagram for the SO$_{2}$ 18$_{4,14}$ -- 18$_{3,15}$ line, employing a PA of 157$\degr$, with the peak emission of each channel superimposed. The peak emission was obtained through the CASA task \texttt{imfit} and the offset position was calculated by projecting the peak emission onto the disk position angle. The redshifted emission is more extended than the blueshifted emission:  the former shows emission up to 0$\farcs$35 ($\sim$50~au) and the latter up to 0$\farcs$21 ($\sim$30~au). The central and right panels of Fig.~\ref{fig:pv_25} are  zoomed-in versions of the red- and blueshifted emission. The high-velocity points are best fitted with an infalling-rotating profile (\textit{V$_\mathrm{rot+inf}$}), employing the equation

\begin{equation} 
    V_\mathrm{rot+inf} = \frac{\sqrt{2 G M_{\star} sin(i) r_{CB}}}{r} \ , 
    \label{eq:Eq2}
\end{equation}

\noindent where \textit{G} is the gravitational constant, \textit{M$_{\star}$} the protostellar mass, \textit{r$_\mathrm{CB}$} the radius of the centrifugal barrier, \textit{i} the inclination of the disk, and \textit{r} the distance from the protostar. Equation \ref{eq:Eq2} is  from \cite{Oya2014}, and the inclination term has been added explicitly. The \textit{r$_\mathrm{CB}$} is given by the maximum radial velocity \citep{Sakai2014} and can be estimated from the PV diagram; the maximum radial velocity of 17.5~km~s$^{-1}$ corresponds to \textit{r$_\mathrm{CB}$}~=~ 0$\farcs$08 ($\sim$11~au). The maximum radial velocity  changes  depending on the assumption of \textit{V$_\mathrm{sys}$}; therefore, if \textit{V$_\mathrm{sys}$} changes by 0.5~km~s$^{-1}$, \textit{r$_\mathrm{CB}$} will change by 0$\farcs$01 ($\sim$1.4~au). This leaves us with a degeneracy in the protostellar mass and the inclination, given by the term \textit{M$_{\star}$}sin(\textit{i}). A protostellar mass of 1.5~M$_{\odot}$ is obtained if an inclination value of 70$\degr$ is assumed, following the interpretation of \cite{Terebey1992} that the outflow axis of IRS~44 lies close to the plane of the sky (from VLA observation of water masers). If we use other inclination values, such as 50$\degr$ and 90$\degr$, the points are well fitted with an infalling-rotating profile with a \textit{M$_{\star}$} of 1.8~M$_{\odot}$ and 1.4~M$_{\odot}$, respectively. \cite{Seifried2016} proposed that the protostellar mass can be estimated by fitting the maximum velocity offset in the PV diagram (i.e., the borders above 3$\sigma$, which correspond to the outer envelope), instead of fitting the peak emission of each channel. Following this procedure, a protostellar mass of 4~M$_{\odot}$ is obtained. 

The centrifugal barrier is the radius at which most of the gas kinetic energy contained in infalling motion is converted to rotational motion. The gas motion of the disk--envelope system outside \textit{r$_\mathrm{CB}$} can be regarded as infalling-rotating motion, while that inside can be regarded as Keplerian motion \citep{Sakai2014, Oya2018}. The \textit{r$_\mathrm{CB}$} is half of the centrifugal radius (\textit{r$_\mathrm{CR}$}), beyond which the gas is falling \citep{Oya2018}. From the SO$_{2}$ 18$_{4,14}$ -- 18$_{3,15}$ line, \textit{r$_\mathrm{CB}$}~=~0$\farcs$08 ($\sim$11~au) and \textit{r$_\mathrm{CR}$}~=~0$\farcs$16 ($\sim$22~au). Beyond \textit{r$_\mathrm{CR}$} the more extended redshifted emission is seen, while no blueshifted counterpart is observed. This is consistent with the redshifted protrusion seen in the contour maps of Fig.~\ref{fig:contour} at velocities between 6 and 10~km~s$^{-1}$, suggesting that a localized streamer might be infalling toward the system and, when entering the centrifugal radius at 0$\farcs$16, an infalling-rotating profile dominates the dynamics. A Keplerian disk is expected inside the centrifugal barrier of 0$\farcs$08; however, this is close to the resolution of our data and the presence of a Keplerian disk is not conclusive with the current data. If a Keplerian disk exists toward IRS~44, its radius will be $\leq$0$\farcs$08 ($\sim$11~au). Given that no Keplerian motions are observed in our data, the rotational signature seen in the moment 1 map of SO$_{2}$ (Fig.~\ref{fig:central})  suggests the presence of a disk--envelope structure and not a rotationally supported disk.

\subsection{Column densities, kinetic temperatures, and optical depth}

In this section we estimate kinetic temperatures (\textit{T$_\mathrm{kin}$}), SO$_{2}$ and SO molecular column densities (\textit{N$_\mathrm{SO_2}$} and \textit{N$_\mathrm{SO}$}), and the optical depth of the lines by employing the non-LTE radiative transfer code RADEX \citep{vanderTak2007}. Later on, rotational temperatures (\textit{T$_\mathrm{rot}$}) and \textit{N$_\mathrm{SO_2}$} of optically thin lines are estimated from the rotational diagram method, and excitation temperatures (\textit{T$_\mathrm{ex}$}) are assessed from optically thick lines.

\subsubsection{Radiative transfer}

The six different SO$_{2}$ transitions were employed to derive the gas density and temperature by comparing the observed relative intensities with those predicted by RADEX. The observed relative intensities are the quotient between the  moment 0 maps, which present emission up to a radius of $\sim$0$\farcs$2. RADEX was run for a set of kinetic temperatures from 30 to 300~K, SO$_{2}$ column densities from 10$^{12}$ to 10$^{18}$~cm$^{-2}$, and H$_{2}$ number density \textit{n$_\mathrm{H}$} between 10$^{3}$ and 10$^{9}$~cm$^{-3}$. Collisional rates for SO$_{2}$ were taken from the Leiden atomic and molecular database \citep[LAMDA; ][]{Balanca2016}. A value of 5~km~s$^{-1}$ was   used for the broadening parameter (\textit{b}), which corresponds to the line width observed in pixels far from the SO$_{2}$ peak. The brightest SO$_{2}$ line, which is associated with an \textit{E$_\mathrm{up}$} of 197~K, is used as a reference line.

\begin{figure*}[h]
        \centering
        \includegraphics[width=.33\textwidth]{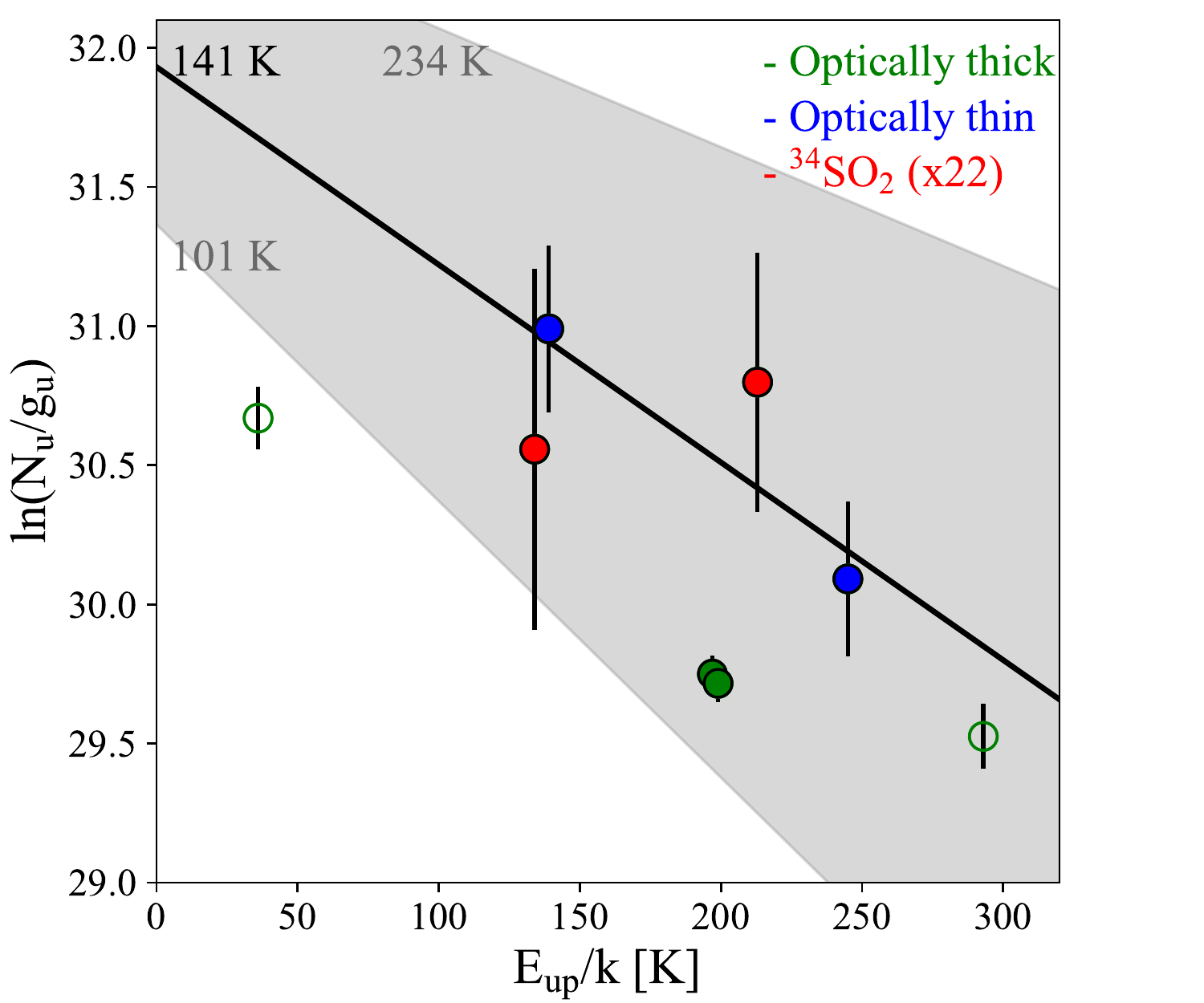}
        \hspace*{-0.4cm}
        \includegraphics[width=.33\textwidth]{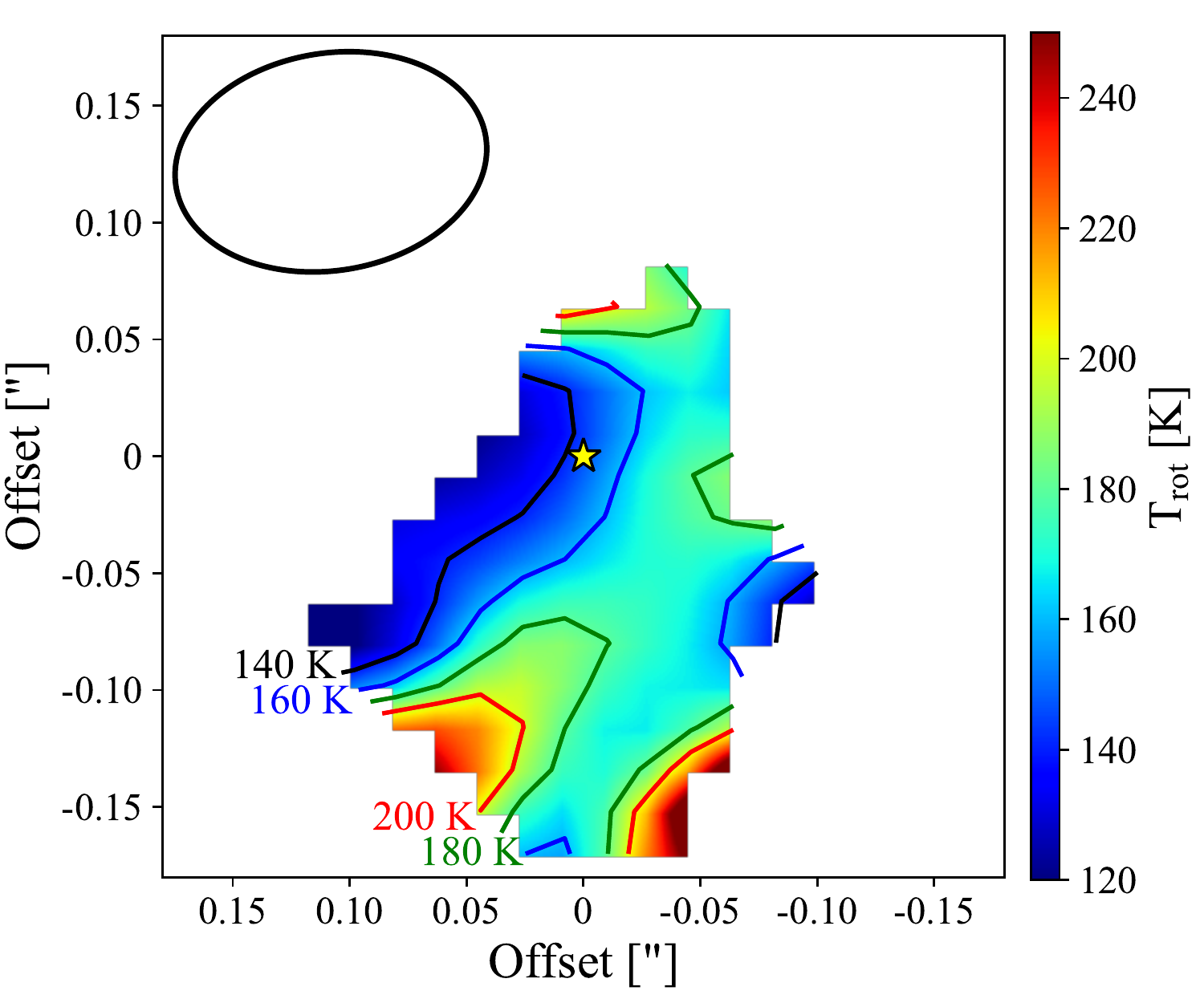}
        \hspace*{0.2cm}
        \includegraphics[width=.33\textwidth]{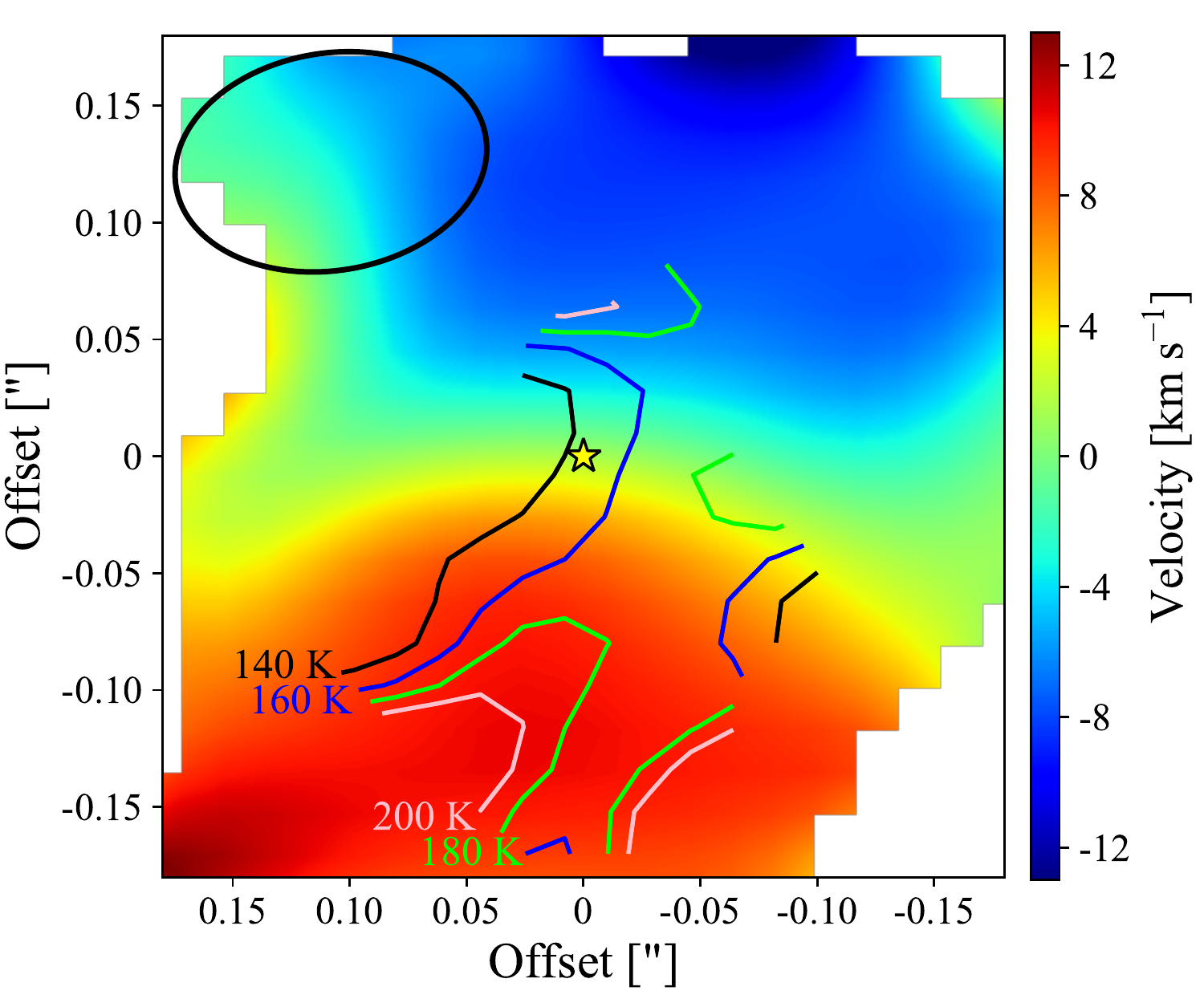}
        \caption[]{\label{fig:Trot}
        Temperature structure of IRS~44. \textit{Left}: Rotational diagram at the source position, where only the optically thin SO$_{2}$ transitions (blue dots) and the $^{34}$SO$_{2}$ lines (red dots) are used for the fit. The abundance ratio $^{32}$S/$^{34}$S~=~22 is  from \cite{Wilson1999}. Optically thick SO$_{2}$ transitions (green dots) and those lines where the optical depth is not conclusive (open dots) show a significant offset with respect to the other lines, and they were not included in the calculation of the rotational temperature and SO$_{2}$ column densities. \textit{Center}: Rotational temperature map created from optically thin SO$_{2}$ and $^{34}$SO$_{2}$ transitions. Contours show specific values of 140, 160, 180, and 200~K. \textit{Right}: Moment 1 map (right panel of Fig.~\ref{fig:central}) with the same specific values of the rotational temperature as those of the central panel, showing that the highest temperatures coincide with the redshifted protrusion. The synthesized beam is shown by the black ellipse in the upper left corner and the yellow star indicates the position of the source.
         }
\end{figure*}

RADEX models with \textit{n$_\mathrm{H}$} between 10$^{3}$ and 10$^{7}$~cm$^{-3}$ were   unable to explain the observed line ratios. The observed relative intensities are shown in Fig.~\ref{fig:ratio}, and they are compared with RADEX results for a H$_{2}$ number density of 10$^{9}$ and 10$^{8}$~cm$^{-3}$. The observed values provide a range of possibilities for \textit{T$_\mathrm{kin}$} and \textit{N$_\mathrm{SO_2}$}, given \textit{n$_\mathrm{H}$}. For \textit{n$_\mathrm{H}$}~=~10$^{8}$~cm$^{-3}$, there are no possible values that satisfy all the observed ranges, implying that the SO$_{2}$ emitting region is associated with \textit{n$_\mathrm{H}$}~$\textgreater$~10$^{8}$~cm$^{-3}$. On the other hand, for \textit{n$_\mathrm{H}$}~=~10$^{9}$~cm$^{-3}$, \textit{T$_\mathrm{kin}$} should be higher than 90~K and 8~$\times$~10$^{16}$~$\leq$~\textit{N$_\mathrm{SO_2}$}~$\leq$~8~$\times$~10$^{17}$~cm$^{-2}$. This possible values are shown in Fig.~\ref{fig:radex}.  

For \textit{n$_\mathrm{H}$}~=~10$^{9}$~cm$^{-3}$, the optical depth of the six SO$_{2}$ lines is analyzed, taking into account the possible values of \textit{T$_\mathrm{kin}$} and \textit{N$_\mathrm{SO_2}$}. Figure~\ref{fig:tau} shows that, from the six SO$_{2}$ lines, two are optically thick (SO$_{2}$ 18$_{4,14}$ -- 18$_{3,15}$ and SO$_{2}$ 20$_{1,19}$ -- 19$_{2,18}$), two are optically thin (SO$_{2}$ 16$_{7,9}$ -- 17$_{6,12}$ and SO$_{2}$ 10$_{6,4}$ -- 11$_{5,7}$), and nothing conclusive can be said about the remaining two (SO$_{2}$ 5$_{3,3}$ -- 4$_{2,2}$ and SO$_{2}$ 24$_{2,22}$ -- 23$_{3,21}$).

\subsubsection{Rotational diagram}

For optically thin SO$_{2}$ lines and the less abundant isotopolog $^{34}$SO$_{2}$, the beam-averaged column densities and rotational temperatures can be assessed by the rotational diagram analysis, summarized by \cite{Goldsmith1999}. The gas is assumed to be under local thermodynamic equilibrium (LTE); therefore, all the molecular transitions can be characterized by a single excitation temperature, also called rotational temperature (\textit{T$_\mathrm{rot}$}). In this regime the following equation is valid:

\begin{equation} 
    \ln \frac{N_\mathrm{u}}{g_\mathrm{u}} = \frac{-1}{T_\mathrm{rot}} \frac{E_\mathrm{u}}{k} + \ln \frac{N}{Q(T_\mathrm{rot})} \ .
    \label{eq:Eq4}
\end{equation}

\noindent Here \textit{N$_\mathrm{u}$} is the column density of the upper level, \textit{g$_\mathrm{u}$} the level degeneracy, \textit{E$_\mathrm{u}$}/\textit{k} the energy of the upper level in K, \textit{k} the Boltzmann constant, \textit{N} the total column density of the molecule, and \textit{Q(T$_\mathrm{rot}$)} the partition function that depends on the rotational temperature. 

Under the optically thin condition, \textit{N$_\mathrm{u}$} is obtained from

\begin{equation} 
    N_\mathrm{u} = \frac{8 \pi k \nu^{2} W}{h c^{3} A_{ul}}             \ ,
    \label{eq:Eq5}
\end{equation}

\noindent where $\nu$ is the line frequency, \textit{W} the integrated line intensity, \textit{c} the speed of light, and \textit{A$_\mathrm{ul}$} the Einstein coefficient for spontaneous emission. Equation~\ref{eq:Eq5} can be rewritten as

\begin{equation} 
    N_\mathrm{u} = 1943.59 \ \left( \frac{\nu}{1 \ \mathrm{GHz}}\right)^{2} \ \left(\frac{W}{1 \ \mathrm{K \ km \ s^{-1}}}\right) \ \left(\frac{1 \ \mathrm{s^{-1}}}{A_\mathrm{ul}}\right)                 \ ,
    \label{eq:Eq6}
\end{equation}

\noindent where \textit{N$_\mathrm{u}$} is  obtained in units of cm$^{-2}$.

\begin{figure*}[h]
        \centering
        \includegraphics[width=.98\textwidth]{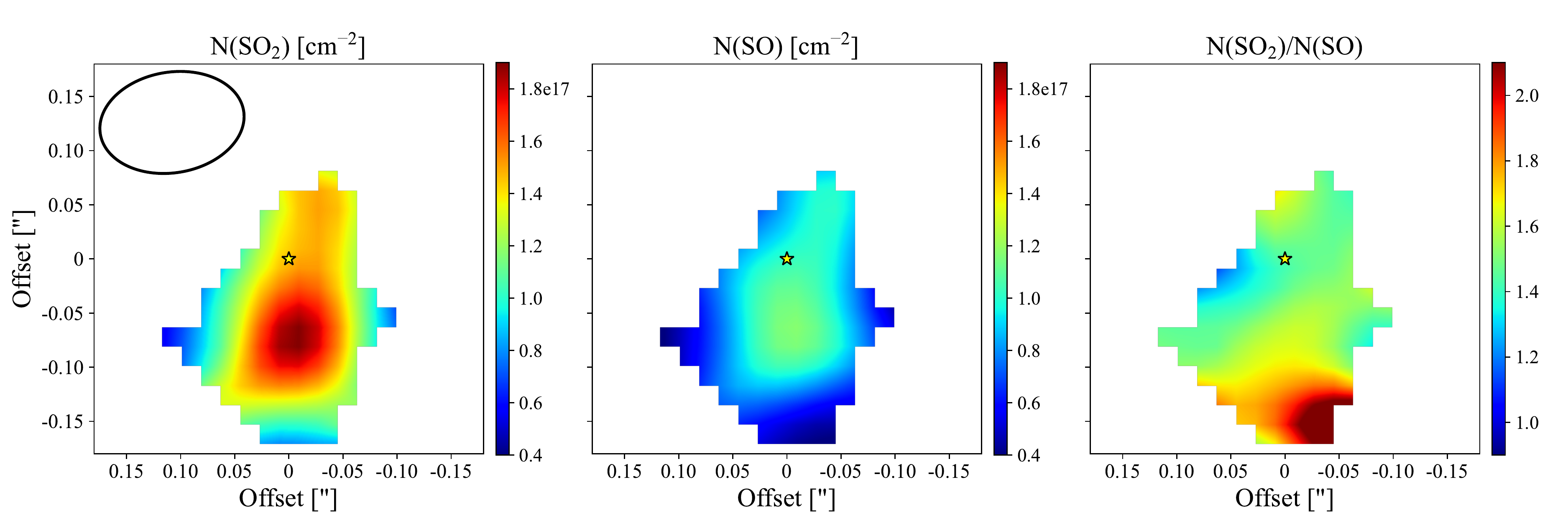}
        \caption[]{\label{fig:column_density}
        Column densities. \textit{Left}: SO$_{2}$ column density obtained from the rotational diagram method. The synthesized beam is shown by the black ellipse in the upper left corner. \textit{Center}: SO column density, assuming optically thin emission and employing the rotational temperatures from Fig.~\ref{fig:Trot}. \textit{Right}: Column density ratio of SO$_{2}$ to SO. The yellow star indicates the position of the source.
         }
\end{figure*}

Equations~\ref{eq:Eq4} and~\ref{eq:Eq6} were used to calculate \textit{T$_\mathrm{rot}$} and create the map shown in Fig.~\ref{fig:Trot}. For each pixel, only the optically thin SO$_{2}$ lines and $^{34}$SO$_{2}$ isotopologs were used to fit the rotational temperature. For $^{34}$SO$_{2}$, an abundance ratio $^{32}$S/$^{34}$S~=~22 \citep{Wilson1999} was adopted. The left panel of Fig.~\ref{fig:Trot} shows the example of the fit from the pixel that corresponds to the source position and a clear offset is seen between optically thin (blue and red dots) and optically thick lines (green dots). The detection of optically thin lines and the less abundant isotopolog, $^{34}$SO$_{2}$, is crucial for an accurate estimate of the rotational temperature, and consequently for the SO$_{2}$ column density as well. The \textit{T$_\mathrm{rot}$} map (central and right panels of Fig.~\ref{fig:Trot}) shows high temperatures ($\geq$120~K) in the region where the SO$_{2}$ emission arises. In addition, the warmest region, southeast from the protostar, seems to correlate with the redshifted protrusion. When infalling material reaches the surface layers of the disk--envelope structure, it generates accretion shocks that are predicted to increase the temperature and the density by up to two orders of magnitude \citep[$\sim$10$^{9}$~cm$^{-3}$;][]{vanGelder2021}. If the dust temperature exceeds 60~K, SO$_{2}$ molecules can efficiently desorb from dust grains. 

Figure~\ref{fig:column_density} shows the SO$_{2}$ and SO column densities, and the ratio between them. The region where the six SO$_{2}$ lines are detected shows \textit{N$_\mathrm{SO_{2}}$} values between 1.0 and 1.8~$\times$~10$^{17}$~cm$^{-2}$, while \textit{N$_\mathrm{SO}$} presents lower values, between 0.6 and 1.3~$\times$~10$^{17}$~cm$^{-2}$. Since there is only one observed (and detected) SO line, RADEX was employed using the same temperature and density parameters as SO$_{2}$ (i.e., \textit{n$_\mathrm{H}$}~=~10$^{9}$~cm$^{-3}$ and \textit{T$_\mathrm{kin}$}~=~90~K), concluding that this SO line in particular is optically thin. The column density ratio between SO$_{2}$ and SO is shown in the right panel of Fig.~\ref{fig:column_density} and it is found to be higher than 1 toward the SO$_{2}$ emitting region.

\subsubsection{Optically thick lines}

As seen in  Sect. 4.2.1, two out of six SO$_{2}$ lines are optically thick. For optically thick lines, the peak temperature (\textit{T$_\mathrm{peak}$}) provides a good measure of (\textit{T$_\mathrm{ex}$}) with

\begin{equation} 
    T_\mathrm{ex} = \frac{E_\mathrm{up}}{k} \left[ \mathrm{log} \left( \frac{E_\mathrm{up}}{kT_\mathrm{peak}} + 1 \right) \right]^{-1}                \ .
    \label{eq:Eq7}
\end{equation}

\begin{figure}[h]
        \centering
        \includegraphics[width=0.48\textwidth]{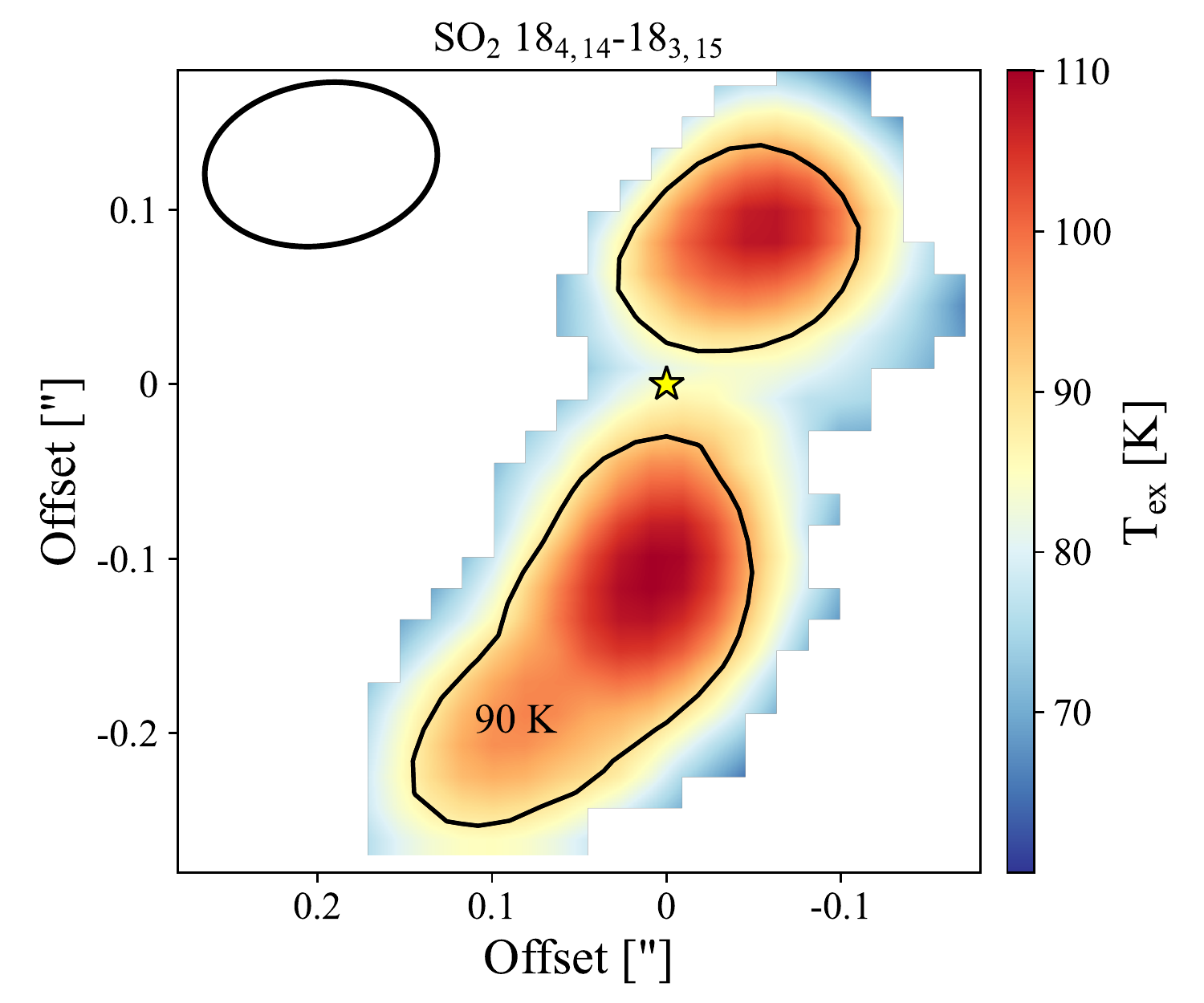}
        \caption[]{\label{fig:Tex}
        Excitation temperature of the optically thick SO$_{2}$ 18$_{4,14}$--18$_{3,15}$ transition using Eq.~\ref{eq:Eq7}. The black contour represents the specific value of 90~K. The yellow star indicates the position of the source and the synthesized beam is shown by the black ellipse in the upper left corner.
                 }
\end{figure}

\noindent Equation~\ref{eq:Eq7} is    from \cite{Goicoechea2016} and, if \textit{n$_\mathrm{H}$} is much higher than the critical density of the transition (\textit{n$_\mathrm{crit}$}), the line is close to thermalization and \textit{T$_\mathrm{ex}$} approaches \textit{T$_\mathrm{gas}$}. From the two optically thick SO$_{2}$ lines, the line with the lowest \textit{n$_\mathrm{crit}$} ($\sim$10$^{7}$~cm$^{-3}$) corresponds to SO$_{2}$ 18$_{4,14}$--18$_{3,15}$. This transition line was used to create the temperature map shown in Fig.~\ref{fig:Tex}, where \textit{T$_\mathrm{peak}$} was obtained from a moment 8 map (which provides the maximum value of the spectrum in each pixel). The southern region presents a more extended and elongated structure in the excitation temperature map, consistent with the redshifted protrusion (see also Figs.~\ref{fig:central}, \ref{fig:contour}, and \ref{fig:Trot}), and \textit{T$_\mathrm{ex}$}~$\geq$~70~K are found for the SO$_{2}$ emitting region. Given that the $\tau$ value of this line lies between 1 and 7 (see first panel of Fig.~\ref{fig:tau}), it may not be fully thermalized, and therefore the temperature map in Fig.~\ref{fig:Tex} represents a lower limit for \textit{T$_\mathrm{ex}$}. These excitation temperatures are consistent with those found in Sect.~4.2.2 from the rotational diagram method using optically thin transitions.

\section{Discussion}

\subsection{Accretion shocks, disk winds, or outflows?}

The molecules SO and SO$_{2}$ are known as shock tracers and there are three main physical origins for these shocks: outflows \citep[e.g.,][]{Tafalla2010, Persson2012}, disk winds \citep[e.g.,][]{Tabone2017}, and accretion shocks \citep[e.g.,][]{Sakai2014, Garufi2022}. 

For IRS 44 the outflow scenario can be ruled out from the shape of the PV diagram shown in Fig.~\ref{fig:pv_25} and the high densities ($\geq$10$^{8}$~cm$^{-3}$) found for the SO$_{2}$ emitting region. PV diagrams related with outflow emission show that the velocity linearly increases as a function of the distance to the protostar \citep[e.g.,][]{Lee2000,Arce2013} and densities below 10$^{8}$~cm$^{-3}$ have been found in the inner regions of the outflow cavity associated with young protostars \citep{Kristensen2013}. In addition, the broadness of the SO$_{2}$ lines rules out the envelope origin, where typical line widths are below 2~km~s$^{-1}$ \citep[e.g.,][]{Harsono2021}. 

As disk winds are related with gas that is ejected at small radial distances from the central source \citep[e.g.,][]{Bjerkeli2016, Alves2017}, some degree of symmetry is expected on the surface layers of the disk, such as a butterfly shape. \cite{Tabone2017} have proposed that the SO and SO$_{2}$ emission detected toward the Class 0 source HH212 originates from a disk wind between $\sim$50 and $\sim$150~au. Nevertheless, \cite{Panoglou2012} have shown that species such as SO survive between 10 and 100~au in disk winds toward Class 0 sources, but they get destroyed by photodissociation beyond $\sim$1~au in disk winds from more evolved Class I sources. The SO$_{2}$ emission does not show the expected symmetry for a disk wind and the kinematic analysis indicates that the material follows an infalling-rotating profile without a Keplerian signature. If disk winds are present, we expect them to arise from the disk surface layers, likely inside 0$\farcs$08 (11~au). 

The high temperatures estimated from optically thin ($\geq$120~K) and optically thick ($\geq$70~K) lines (Figs.~\ref{fig:Trot} and \ref{fig:Tex}), the moderate velocities (between 12 and 14~km~s$^{-1}$), and the high densities ($\geq$~10$^{8}$~cm$^{-3}$) found for IRS 44 are in agreement with the accretion shock scenario. \cite{vanGelder2021} have shown that accretion shocks can efficiently desorb SO$_{2}$ from dust grains when moderate velocities ($\geq$~10~km~s$^{-1}$) and high densities ($\geq$~10$^{8}$~cm$^{-3}$) are present. For densities above 3~$\times$~10$^{4}$~cm$^{-3}$, the gas and the dust are efficiently coupled, \textit{T$_\mathrm{dust}$}~=~\textit{T$_\mathrm{gas}$} \citep{Evans2001, Galli2002}, and a dust temperature above 62~K is required in order to sublimate SO$_{2}$ molecules from dust grains \citep{Penteado2017,vanGelder2021}. In interstellar ices, SO$_{2}$ is tentatively detected \citep{Boogert1997, Zasowski2009}; however, chemical models predict that SO$_{2}$ is the most abundant species in the gas in the warm-up phase, when the protostar is formed \citep{Woods2015}. 

Accretion shocks would also desorb SO molecules form dust grains and the gas-phase abundance of SO$_{2}$ could increase through the reaction of SO with OH \citep{Charnley1997, vanGelder2021}. Nevertheless, \cite{Karska2018} did not detect OH toward IRS 44 from \textit{Herschel}/PACS observations, suggesting that the gas-phase formation of SO$_{2}$ by oxidation of SO could be ruled out. SO$_{2}$/SO~$\geq$~1 also suggests that the radiation field from the protostar is not efficiently photo-dissociating SO$_{2}$ into SO \citep[e.g.,][]{Booth2021} and that the cosmic ray ionization rate is low \citep[$\zeta$=~1.3~$\times$~10$^{17}$~s$^{-1}$,][]{Woods2015}. 

In this section we suggest that SO and SO$_{2}$ molecules toward IRS 44   sublimate from heated dust grains by the accretion shocks with moderate velocity shocks ($\geq$~10~km~s$^{-1}$) and high densities ($\geq$~10$^{8}$~cm$^{-3}$). If there is a chemical reaction that contributes to the SO$_{2}$ abundance in the gas phase, it should be a different one from the reaction of SO with OH. Future observations of other molecular species, such as OCS, H$_{2}$S, and H$_{2}$CO, will confirm the formation path of SO$_{2}$:  direct desorption from dust grains, gas-phase formation, or a combination of both. H$_{2}$S and H$_{2}$CO are directly linked to the gas-phase formation of SO and SO$_{2}$, while OCS presents a similar desorption temperature to  SO, but it does not participate in the gas-phase chemistry \citep{Charnley1997}.

\subsection{Morphology of IRS~44}

\begin{figure*}[h]
        \centering
        \includegraphics[width=0.7\textwidth]{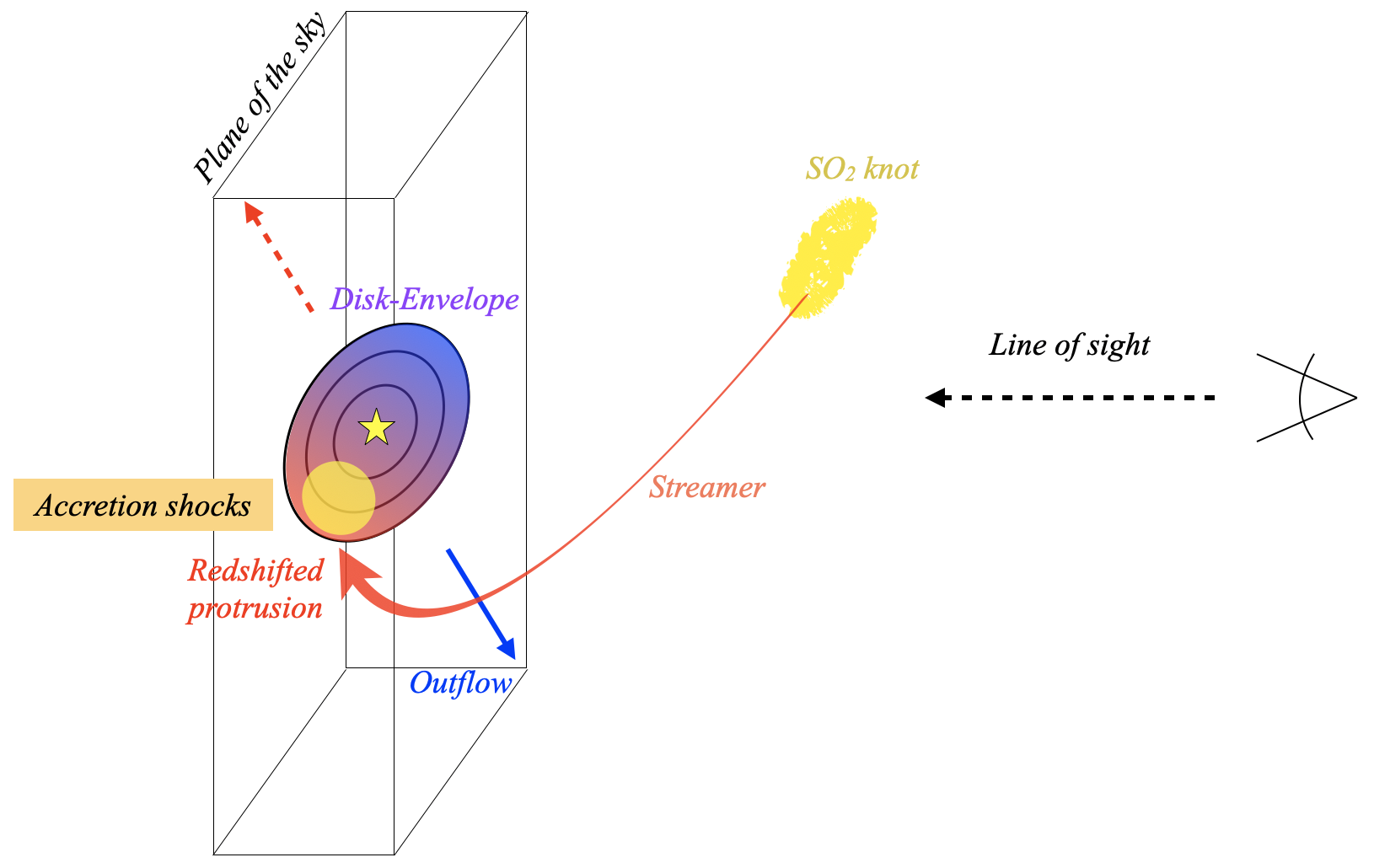}
        \caption[]{\label{fig:cartoon}
        Schematic representation of IRS~44. The quiescent SO$_{2}$ emission (knot) would be part of a streamer that allows material to fall into the disk--envelope through the redshifted protrusion and generate accretion shocks. These shocks sublimate SO$_{2}$ molecules from dust grains and enhance the SO$_{2}$ gaseous abundance, showing an emission peak $\sim$0$\farcs$1 to the south of the protostar. 
                 }
\end{figure*}

Given that \textit{(i)} quiescent and colder SO$_{2}$ emission is present at $\sim$2$\farcs$8, \textit{(ii)} a redshifted protrusion is seen at velocities between 2 and 10~km~s$^{-1}$, \textit{(iii)} the highest temperatures seem to correlate with the redshifted protrusion, \textit{(iv)} blueshifted material beyond 0$\farcs$2 ($\sim$30~au) is absent,  and \textit{(v)} the SO$_{2}$ emission peak is observed at a distance of $\sim$0$\farcs$1 from the continuum peak (at redshifted velocities), a localized streamer might be accreting material to the envelope--disk system and generating accretion shocks that release SO$_{2}$ molecules from the dust to the gas phase. Figure~\ref{fig:cartoon} shows a schematic representation of IRS~44 and the localized streamer, which would be located between the observer and the disk--envelope component and would feed the system toward the redshifted protrusion. As IRS 44 is classified as a Class I source, meaning that the envelope is still present but largely dissipated \citep[\textit{M$_\mathrm{env}$}~=~0.051~M$_{\odot}$ for a distance of 139~pc;][]{Jorgensen2008, Jorgensen2009}, it is more likely that the infalling of material occurs through streamers and not in a spherically symmetric way. A similar behavior is seen toward the Class I source TMC-1A, as asymmetric CS and SO emission is explained by a cloudlet capture and subsequent formation of an infalling streamer \citep{Hanawa2022}, and the more evolved Class I/II sources DG Tau and HL Tau \citep{Garufi2022}, where accretion shocks traced by SO and SO$_{2}$ are located along the late infalling streamers still feeding the system.

\subsection{Outflow direction vs. disk--envelope direction}

As seen in Fig.~\ref{fig:central}, the disk--envelope direction (157$\degr$) is not perpendicular to that of the single-dish outflow (20$\degr$). The  latter is seen at large scales ($\sim$30$\arcsec$) and the outflow direction may vary due to the surrounding gas. If the misalignment is real, this could be due to the presence of a binary component, a cloudlet capture, or physical processes during the formation history, such as misalignment between the cloud rotation axis and the initial B-field direction, formation from a turbulent core, or non-ideal magnetohydrodynamics (MHD) effects. 

\cite{Terebey2001}  proposed that IRS 44 is a protobinary system with a separation of 0$\farcs$27 and PA~=~81$\degr$, based on HST observations. The primary component has been detected at 1.60, 1.87, and 2.05~$\mu$m, while the secondary component is only visible at the longest wavelength, at 2.05~$\mu$m. Nevertheless, there is no sign of binarity toward IRS 44 in the submillimeter regime, following this work with an angular resolution of $\sim$0$\farcs$1 and the ALMA data presented in \cite{Artur2019} and \cite{Sadavoy2019}, which report an angular resolution of $\sim$0$\farcs$4 and $\sim$0$\farcs$25, respectively. The HST emission in 2.05~$\mu$m could therefore be associated with scattered light and not a binary component or the binary could be very faint at submillimeter wavelengths (below our sensitivity). 

In the case of a cloudlet capture, each cloudlet should have a different angular momentum vector and the capture process can potentially change the rotation axis of the disk \citep[e.g., ][]{Dullemond2019, Kuffmeier2020, Hanawa2022}. The presence of a localized streamer toward IRS~44 could be affecting the rotation axis of the disk--envelope, and the result would depend on the mass and the angular momentum vector of the infalling material.

The misalignment between the cloud rotation axis and the initial B-field direction can create a warped disk structure during the protostellar core collapse \citep{Hirano2019}, and B-fields in protostellar cores appear to be randomly aligned with their respective outflows \citep{Hull2014,Lee2017}. In the absence of a binary component, this initial misalignment could explain the change in direction observed toward IRS 44. A similar situation was proposed for the Class I disk L1489 \citep{Sai2020}, where the observation of a warped disk is explained by the initial misaligment between the initial B-field direction and the angular momentum vector. 

Different velocity gradients between the direction of the rotationally supported disk and the direction of the envelope rotation were seen for a handful of Class 0/I sources \citep{Brinch2007b, Harsono2014}. This misalignment might be due to formation from turbulent cores or non-ideal MHD effects, such as the Hall effect \citep[e.g., ][]{Li2011, Braiding2012}.

\subsection{Nondetection of C$^{17}$O and absence of warm CH$_{3}$OH toward IRS~44}

Previous observations of IRS~44 do not detect C$^{17}$O (3--2) and warm CH$_{3}$OH (\textit{E$_\mathrm{up}$}~=~65~K) at an angular resolution of 0$\farcs$4 \citep[$\sim$60~au);][]{Artur2019}. C$^{17}$O is commonly associated with Keplerian disks in Class I sources, and its nondetection might be related with the absence of a Keplerian disk, at least outside 11~au. CH$_{3}$OH, on the other hand, is hardly detected in Class I sources \citep{Artur2019}; however, its gas-phase abundance is enhanced in shocked regions and a correlation between SO$_{2}$ and CH$_{3}$OH is expected. That SO$_{2}$ shows strong emission and CH$_{3}$OH is not detected toward IRS~44 might be related with one of the following possibilities: (\textit{i}) CH$_{3}$OH is being desorbed form dust grains, but later on it is destroyed by the moderate velocities of the shocks \citep[$\geq$10~km~s$^{-1}$;][]{Suutarinen2014}; (\textit{ii}) the formation of CH$_{3}$OH on the grain surfaces, from H$_{2}$CO, is not  efficient; (\textit{iii}) the presence of a disk results in colder gas \citep{Lindberg2014,vanGelder2022}; or (\textit{iv}) optically thick dust can hide the emission of COMs \citep{Nazari2022}. Future observations of H$_{2}$CO could clarify the CH$_{3}$OH nondetection, and clearly there is some uncertainty regarding the origin of the SO$_{2}$ emission. The origin of this may be related to the uncertain carrier of elemental sulfur in protostellar envelopes. This carrier must be subject to destruction in shocks and clearly carry both S and O.

\section{Summary}

This work presents high angular resolution ($\sim$0$\farcs$1, 14~au) ALMA observations of the Class I source IRS 44. The continuum emission at 0.87~mm is analyzed, together with molecular species such as SO, SO$_{2}$, and $^{34}$SO$_{2}$. The main results are summarized below: 

\begin{itemize}
        \item The continuum emission is contained within a radius of $\sim$0$\farcs$2 (30~au) and a total mass (gas + dust) of 4.0~$\times$~10$^{-3}$~M$_{\odot}$ is calculated for IRS 44. Given that no binary component is detected with our sensitivity, an upper limit of 7~$\times$~10$^{-5}$~M$_{\odot}$ is estimated for its total dust mass.
        \item One SO, six SO$_{2}$, and two out of three $^{34}$SO$_{2}$ lines are detected;  all of the detections show two components in their spectra, a blueshifted one and a redshifted one, both with broad linewidths (between -20 and 20~km~s$^{-1}$). At small scales ($<$~0$\farcs$3) the brightest SO$_{2}$ line is associated with high \textit{E$_\mathrm{up}$}~=~197~K, while the SO$_{2}$ line with low \textit{E$_\mathrm{up}$}~=~36~K presents the brightest emission at larger angular scales (between 2$\arcsec$ and 3$\arcsec$), shows narrow lines below 2~km~s$^{-1}$, and has been associated with a shocked region. 
        \item Around the protostar, SO$_{2}$ shows that the redshifted component is more extended than the blueshifted one, likely related with a redshifted protrusion, and the velocity profile is better fitted with an infalling-rotating profile with \textit{M$_{\star}$}~=~1.5~M$_{\odot}$ and \textit{r$_\mathrm{CB}$}~=~0$\farcs$08. The quiescent shocked region and the redshifted protrusion seem to be part of a localized streamer, allowing material to fall to the disk--envelope and generate accretion shocks. No evidence of Keplerian motions are found; however, a Keplerian disk is  expected inside \textit{r$_\mathrm{CB}$}.
        \item The comparison between observed relative intensities of the various lines and RADEX results indicates that the SO$_{2}$ emission around the protostar   arises from a dense region (\textit{n$_\mathrm{H}$}~$\geq$~10$^{8}$~cm$^{-3}$) with kinetic temperatures above 90~K. In addition, two SO$_{2}$ lines are clearly optically thin lines and  two others are optically thick lines. 
        \item The rotational diagram provides kinetic temperatures between 120 and 250~K for the SO$_{2}$ emitting region, where the warmest regions coincide with the location of the redshifted protrusion, and SO$_{2}$ column densities lie between 0.4 and 1.8~$\times$~10$^{17}$~cm$^{-2}$. SO column densities are a little lower, between 0.4 and 1.2~$\times$~10$^{17}$~cm$^{-2}$, and as a consequence the \textit{N}(SO$_{2}$)/\textit{N}(SO$_{2}$) ratio lies between 1.0 and 2.0.
        \item Optically thick SO$_{2}$ lines provide \textit{T$_\mathrm{ex}$} values between 70 and 110~K (regarded as lower limits) and a temperature structure consistent with warmer material arising from the south.  
        \item The high temperatures, compact emission, high \textit{n$_\mathrm{H}$} densities, and moderate velocities agree with the accretion shock scenario, where molecules are being efficiently sublimated from dust grains. We can conclude, therefore, that accretion shocks toward IRS 44 are associated with \textit{n$_\mathrm{H}$}~$\geq$~10$^{8}$~cm$^{-3}$, \textit{T$_\mathrm{kin}$}~$\geq$~90~K, \textit{T$_\mathrm{rot}$} between 120 and 240~K, \textit{T$_\mathrm{ex}$}~$\geq$~70~K, SO$_{2}$ column densities between 0.4 and 1.8~$\times$~10$^{17}$~cm$^{-2}$, and velocities between 12 and 14~km~s$^{-1}$.
        \item Finally, high-energy SO$_{2}$ lines (\textit{E$_\mathrm{up}$}~$\sim$~200~K) seem to be the best tracers of accretion shocks
\end{itemize}

Accretion shocks might have important consequences for the chemical content of the disk and the release of neutral species, such as H$_{2}$O and COMs. It is therefore an important physical process that should be studied in more detail, and high angular resolution observations are essential for this purpose. Future observations of other Class I sources that show bright SO$_{2}$ emission will be necessary to increase the statistics and achieve a more complete picture of accretion shocks. Other molecular species such as CS, OCS, H$_{2}$S, and H$_{2}$CO could provide key additional information. CS is the most abundant sulfur-bearing species in young disks and OCS desorbs from dust grains at a similar temperature than SO, but it does not participate in the gas-phase chemistry below 300~K. H$_{2}$CO and H$_{2}$S are key species in the gas-phase formation of SO and SO$_{2}$. In addition, H$_{2}$CO is a good tracer of the gas temperature and it has similar desorption temperature to SO$_{2}$. Finally, a kinematic study of CO isotopologs, in special C$^{18}$O, could provide information about the existence of a Keplerian disk.

\begin{acknowledgements}

We thank the anonymous referee for a number of good suggestions that helped us to improve this work. This paper makes use of the following ALMA data: ADS/JAO.ALMA$\#$2019.1.00362.S. ALMA is a partnership of ESO (representing its member states), NSF (USA) and NINS (Japan), together with NRC (Canada), MOST and ASIAA (Taiwan), and KASI (Republic of Korea), in cooperation with the Republic of Chile. The Joint ALMA Observatory is operated by ESO, AUI/NRAO and NAOJ. The National Radio Astronomy Observatory is a facility of the National Science Foundation operated under cooperative agreement by Associated Universities, Inc.

E.A.dlV. acknowledges financial support provided by FONDECYT grant 3200797. V.G. acknowledges support from FONDECYT Iniciación 11180904, ANID project Basal AFB-170002, and ANID, -- Millennium Science Initiative Program -- NCN19\_171. J.K.J. acknowledges support from the Independent Research Fund Denmark (grant No. DFF0135-00123B). Daniel Harsono is supported by Centre for Informatics and Computation in Astronomy (CICA) and grant number 110J0353I9 from the Ministry of Education of Taiwan. D.H. acknowledges support from the Ministry of Science of Technology of Taiwan through grant number 111B3005191. N.S. is supported by JSPS KAKENHI grant 20H05845 and pioneering project in RIKEN (Evolution of Matter in the Universe). EvD is supported by the European Research Council (ERC) under the European Union’s Horizon 2020 research and innovation program (grant agreement No. 101019751 MOLDISK).

\end{acknowledgements}

\bibliographystyle{aa} 
\bibliography{References}

\begin{appendix}

\section{SO$_{2}$, $^{34}$SO$_{2}$, and SO detections}

\begin{figure*}[h]
        \centering
        \includegraphics[width=.99\textwidth]{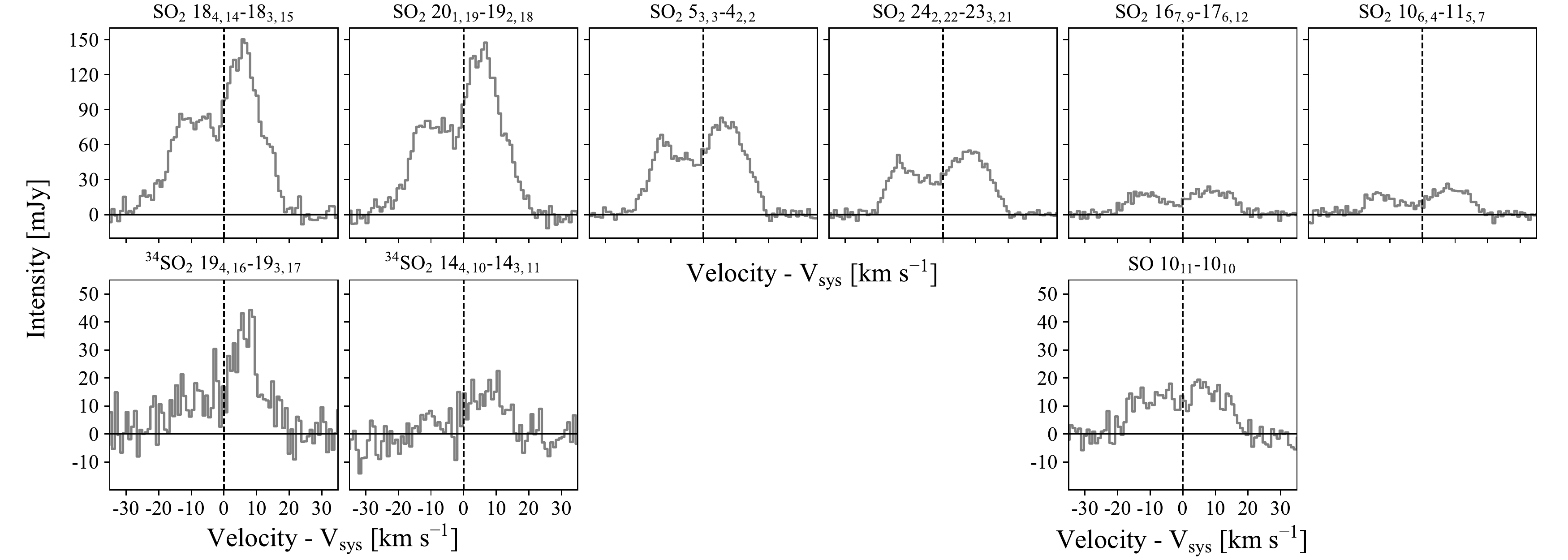}
        \caption[]{\label{fig:spectra}
        Spectra of SO$_{2}$ (\textit{top}), $^{34}$SO$_{2}$ (\textit{bottom left}), and SO (\textit{bottom right}), integrated over a circular region with \textit{r}~=~0$\farcs$2 and centered on the continuum peak position. The systemic velocity corresponds to 3.7~km~s$^{-1}$.
        }
\end{figure*}

\begin{figure*}[h]
        \centering
        \includegraphics[width=.99\textwidth]{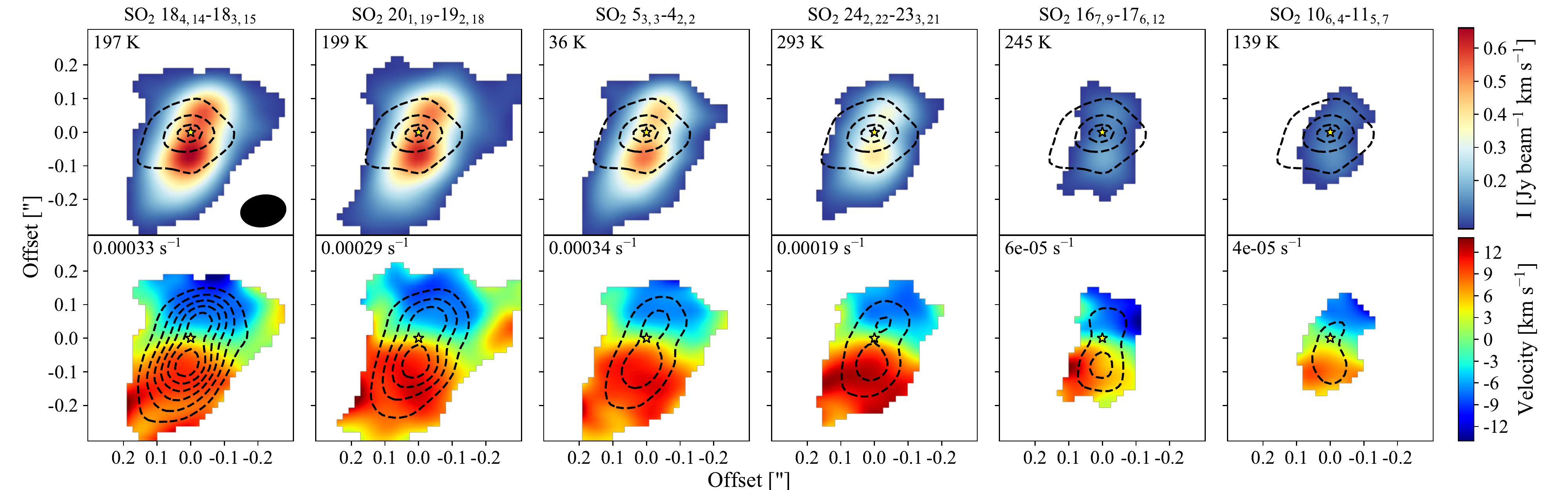}
        \caption[]{\label{fig:mom}
        Small-scale emission. \textit{Top}: Moment 0 maps of SO$_{2}$ above 3$\sigma$ (color scale) and continuum emission (dashed contours). The moment 0 maps were integrated over 60~km~s$^{-1}$ and the continuum contours start at 20$\sigma$ and follow  steps of 80$\sigma$. The \textit{E$_\mathrm{up}$} value of each transition is indicated in the top left corner of each panel and the synthesized beam is shown by the black filled ellipse in the bottom right corner of the first panel. The color scale is the same for the six panels. \textit{Bottom}: Moment 1 maps of SO$_{2}$ above 3$\sigma$ (color scale) and selected values of their respective moment 0 maps (dashed contours). The contours start at 3$\sigma$ and follow  steps of 7$\sigma$, with the exception of the last two panels, which follow  steps  of 3$\sigma$. The \textit{A$_\mathrm{ij}$} value of each transition is indicated in the top left  corner of each panel;  the color scale is the same for the six panels. The adopted systemic velocity is 3.7~km~s$^{-1}$.
        }
\end{figure*}

\begin{figure*}[h]
        \centering
        \includegraphics[width=.7\textwidth]{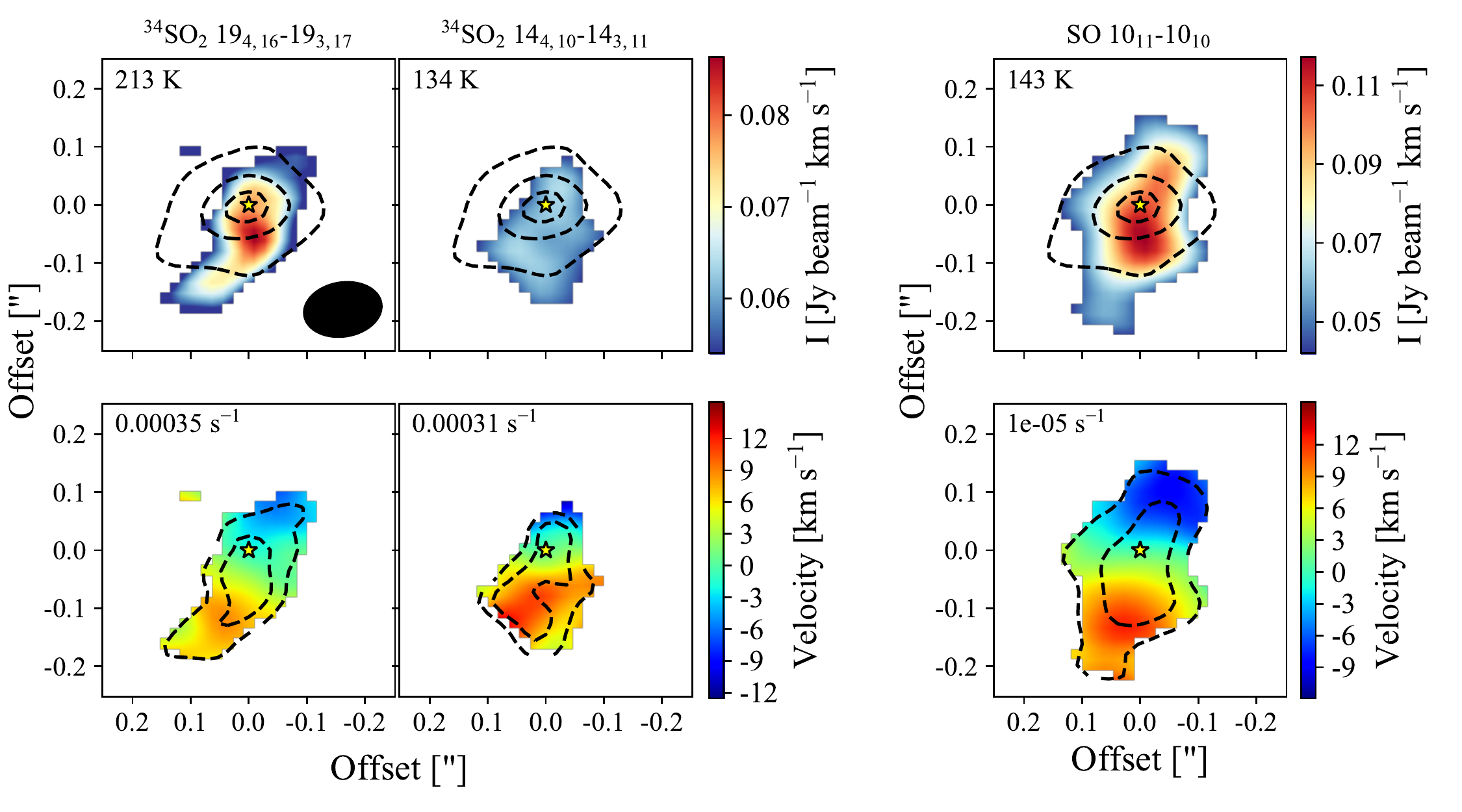}
        \caption[]{\label{fig:mom_bis}
         Same as Fig.~\ref{fig:mom}, but for $^{34}$SO$_{2}$ and SO. The dashed black contours in the moment 1 maps follow  steps of 1$\sigma$ and  3$\sigma$ for $^{34}$SO$_{2}$ and  SO, respectively.
         }
\end{figure*}

\begin{figure*}[h]
        \centering
        \includegraphics[width=.99\textwidth]{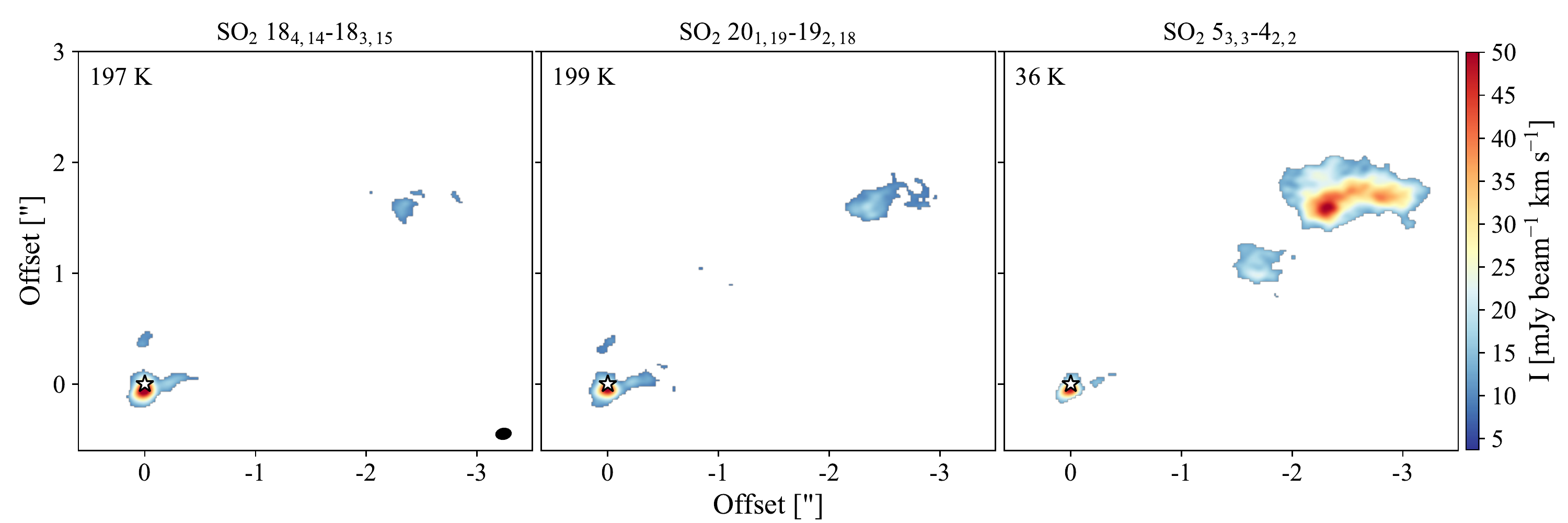}
        \caption[]{\label{fig:mom_large}
        Large-scale emission. Moment 0 maps of SO$_{2}$ integrated over a velocity range of 2~km~s$^{-1}$ and above 1$\sigma$. The white star shows the position of the source and the synthesized beam is indicated by the  black filled ellipse in the  bottom right corner of the first panel. The \textit{E$_\mathrm{up}$} value of each transition is indicated in the top left corner of each panel and the color scale is the same for all three panels.
        }
\end{figure*}

The integrated fluxes of the detected transitions, and upper limits for nondetections, are presented in Table~\ref{table:fluxes}, where a region with \textit{r}~=~0$\farcs$2 and centered on the continuum peak position was chosen. 
The spectra, moment 0, and moment 1 maps of the detected transitions at small scales are shown in Figs.~\ref{fig:spectra}, \ref{fig:mom}, and \ref{fig:mom_bis}, respectively. Figure~\ref{fig:mom_large} presents the large-scale emission, where only three SO$_{2}$ lines were detected.

\begin{table}[ht]
\caption{Integrated fluxes over a velocity range of 60~km~s$^{-1}$ and taking a circular region with \textit{r}~=~0$\farcs$2 centered on the continuum peak position.}
\label{table:fluxes}
\centering
\begin{tabular}{l l r}
	\hline\hline
	Species                         	& Transition					& Flux density                 		\\
 						&                             				& [mJy~km~s$^{-1}$]               	\\
	\hline
	SO                       		& 10$_{11}$ -- 10$_{10}$         		& 624~$\pm$~14          		\\
	SO$_{2}$                        	& 16$_{7,9}$ -- 17$_{6,12}$		& 791~$\pm$~17          		\\
	SO$_{2}$            		& 18$_{4,14}$ -- 18$_{3,15}$  		& 2919~$\pm$~18         		\\
	SO$_{2}$         			& 20$_{1,19}$ -- 19$_{2,18}$  		& 2802~$\pm$~17         		\\
	SO$_{2}$                      	& 24$_{2,22}$ -- 23$_{3,21}$  		& 1782~$\pm$~19         		\\
	SO$_{2}$                  		& 5$_{3,3}$ -- 4$_{2,2}$           		& 2626~$\pm$~23       		\\
	SO$_{2}$                        	& 10$_{6,4}$ -- 11$_{5,7}$ 		& 700~$\pm$~19          		\\
	$^{34}$SO$_{2}$           	& 14$_{4,10}$ -- 14$_{3,11}$  		& 221~$\pm$~13           		\\
	$^{34}$SO$_{2}$         	& 19$_{4,16}$ -- 19$_{3,17}$  		& 547~$\pm$~18            		\\
	$^{34}$SO$_{2}$         	& 9$_{6,4}$ -- 10$_{5,5}$          	& $\leq$~51 $^{(a)}$              	\\
	\hline\hline
\end{tabular}
\tablefoot{$^{(a)}$ 3$\sigma$ upper limit is reported for nondetections.}
\end{table}

\section{Radiative transfer}

\begin{figure*}[h]
        \centering
        \includegraphics[width=.98\textwidth]{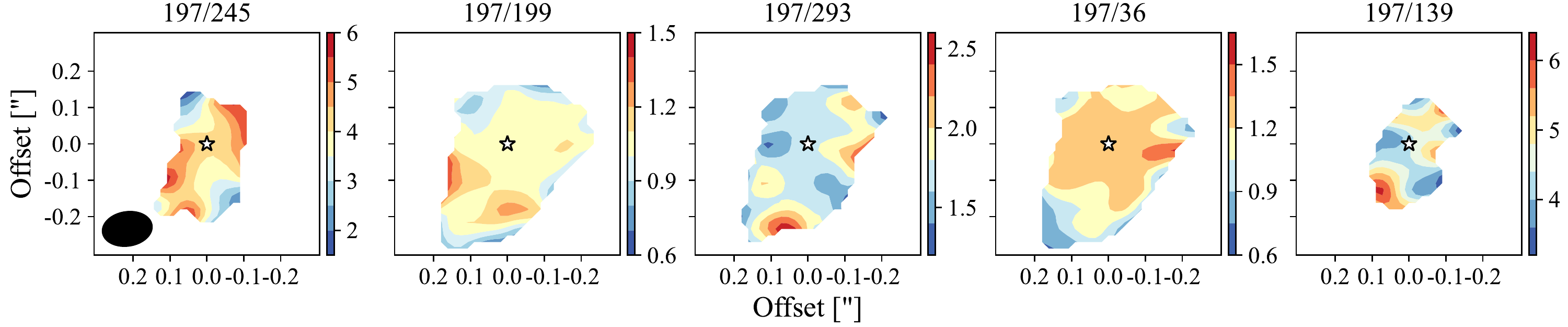}
        \hspace*{0.2cm}
        \includegraphics[width=.99\textwidth]{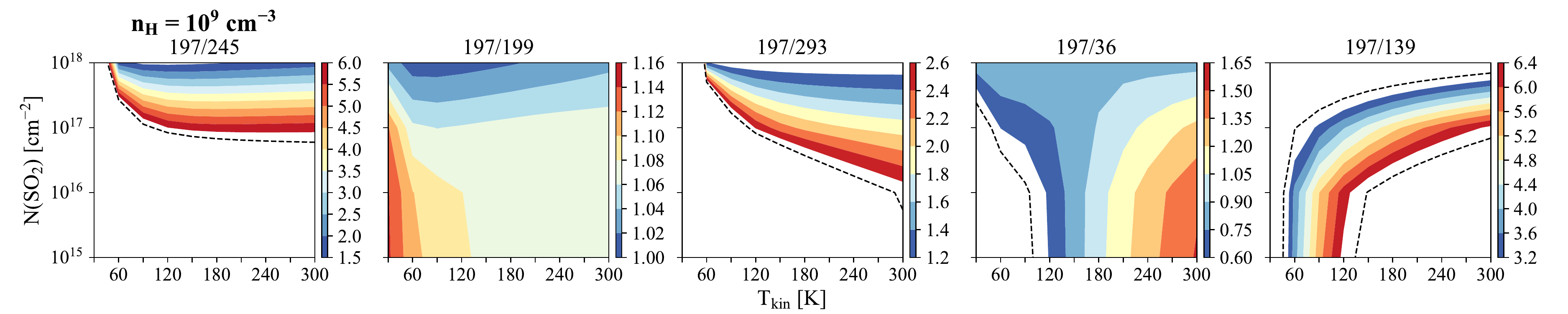}
        \hspace*{0.2cm}
        \includegraphics[width=.99\textwidth]{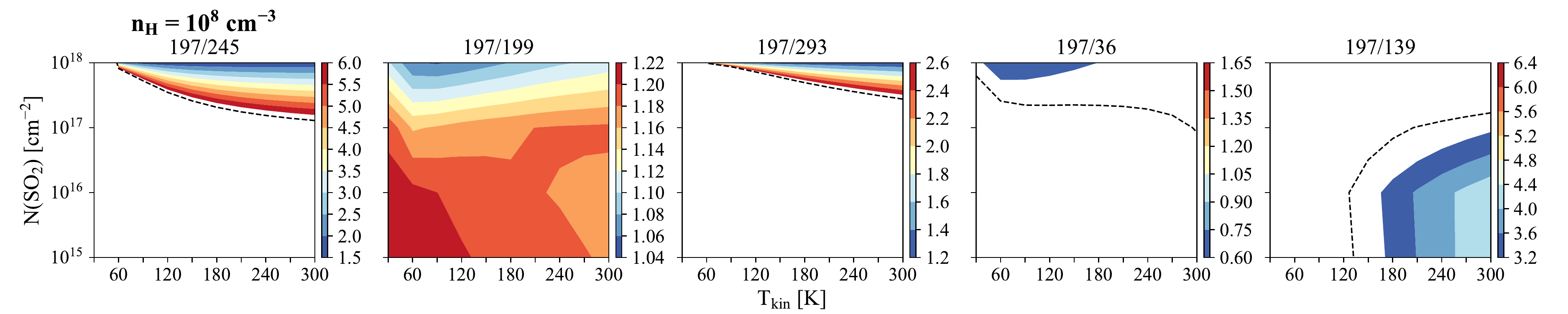}
        \caption[]{\label{fig:ratio}
        Comparison between observed intensity ratios and RADEX models. \textit{Top}: Observed intensity ratios between SO$_{2}$ 18$_{4,14}$--18$_{3,15}$ (\textit{E$_\mathrm{up}$}~=~197~K) and the other five transitions, above a 3$\sigma$ level. The synthesized beam is shown by the black filled ellipse in the  bottom left corner of the first panel and the yellow star indicates the position of the source. \textit{Center}: Intensity ratios from RADEX for the same transitions and employing a H$_{2}$ number density of 10$^{9}$~cm$^{-3}$. \textit{Bottom}: Same analysis from RADEX, but employing a H$_{2}$ number density of 10$^{8}$~cm$^{-3}$. The black dashed contours indicate error values of 1$\sigma$ and the limits of the possible ranges. All values are possible for the panels without black dashed contours.
         }
\end{figure*}

The six different SO$_{2}$ transitions were employed to compare the observed relative intensities with values obtained from RADEX. The observed relative intensities are shown in the upper row of Fig.~\ref{fig:ratio}. The middle and bottom rows show the RADEX results for a H$_{2}$ number density of 10$^{9}$ and 10$^{8}$~cm$^{-3}$, respectively, and the contour levels represent the observed values (shown in the top row). An error value of 1$\sigma$ (employing error propagation) was added to the observed ratios and is represented by the black dashed contours in the RADEX results.  

\begin{figure}[h]
        \centering
        \includegraphics[width=.49\textwidth]{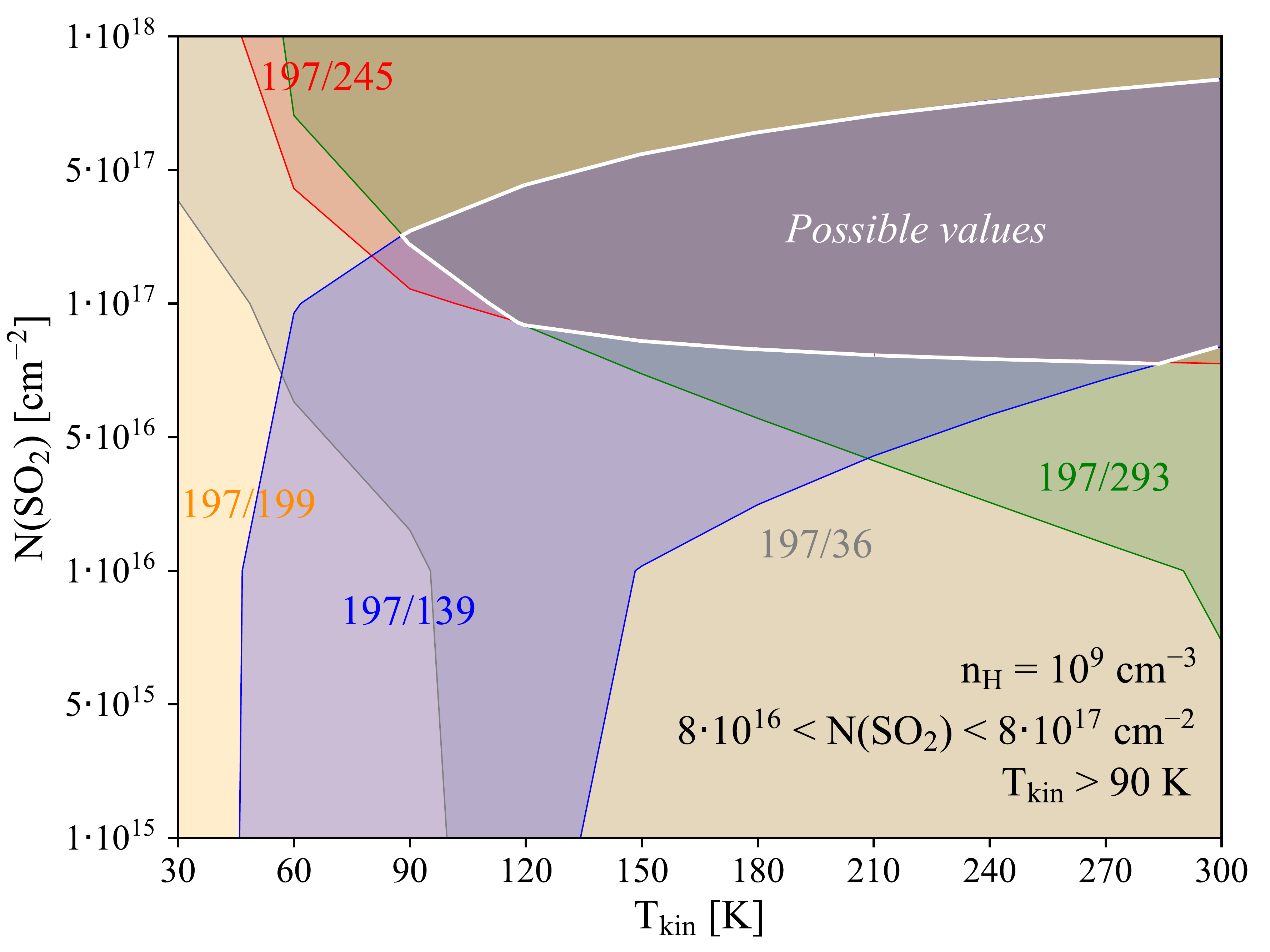}
        \caption[]{\label{fig:radex}
        Range of possible values for the SO$_{2}$ column density and the kinetic temperature (white contours) from the overlap of the observed ranges in Fig.~\ref{fig:ratio} (for a H$_{2}$ number density of 10$^{9}$~cm$^{-3}$), indicated in different colors.
         }
\end{figure}

For \textit{n$_\mathrm{H}$}~=~10$^{8}$~cm$^{-3}$ (bottom row of Fig.~\ref{fig:ratio}), the third panel (197/293) and the fifth panel (197/139) do not present an overlapping region; therefore, this density does not reproduce the observed values and \textit{n$_\mathrm{H}$} should be higher than 10$^{8}$~cm$^{-3}$. On the other hand, for \textit{n$_\mathrm{H}$}~=~10$^{9}$~cm$^{-3}$ (middle row of Fig.~\ref{fig:ratio}), the possible ranges are presented in Fig.~\ref{fig:radex} and the possible values consist of \textit{T$_\mathrm{kin}$}~$\geq$~90~K and  \textit{N$_\mathrm{SO_{2}}$} between 8~$\times$~10$^{16}$ and 8~$\times$~10$^{17}$~cm$^{-2}$. 

\begin{figure*}[h]
        \centering
        \includegraphics[width=1.02\textwidth]{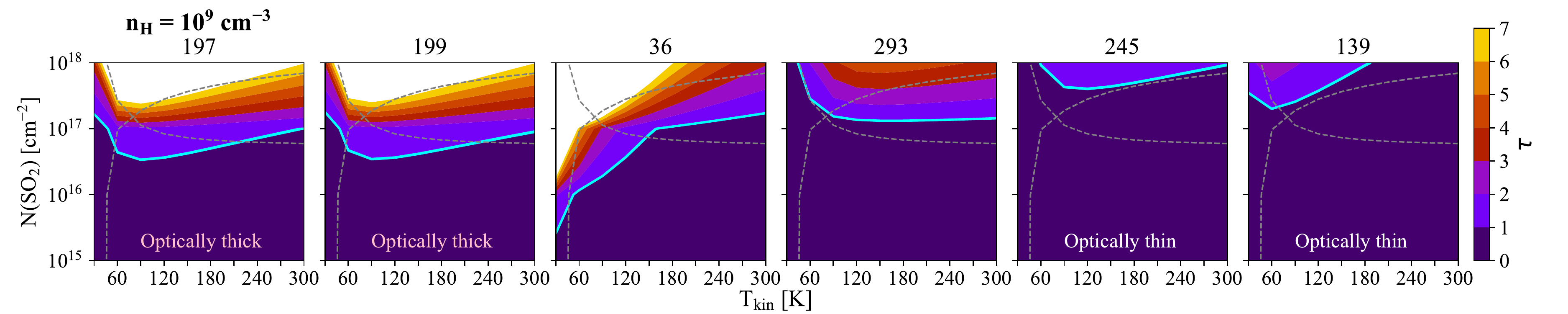}
        \caption[]{\label{fig:tau}
        Optical depth for the six SO$_{2}$ transitions obtained with RADEX with \textit{n$_\mathrm{H}$}~=~10$^{9}$~cm$^{-3}$. The cyan contour represents $\tau$~=~1 and the dashed gray contours indicate the range of possible values for  \textit{N$_\mathrm{SO_{2}}$} and \textit{T$_\mathrm{kin}$} shown in Fig.~\ref{fig:radex}.
         }
\end{figure*}

Figure~\ref{fig:tau} shows the optical depth of the six SO$_{2}$ transitions, obtained with RADEX with \textit{n$_\mathrm{H}$}~=~10$^{9}$~cm$^{-3}$, and the possible values discussed above are shown in gray dashed contours. The brightest transitions, those with \textit{E$_\mathrm{up}$} values of 197 and 199~K, are optically thick lines, while the two weakest ones (\textit{E$_\mathrm{up}$} of 245 and 139~K) are optically thin lines. Nothing conclusive can be said for those transitions with \textit{E$_\mathrm{up}$} values of 36 and 293~K.

\end{appendix}

\end{document}